\documentclass[%
 reprint,
 amsmath,amssymb,
 aps,
]{revtex4-1}

\usepackage{graphicx}
\usepackage{dcolumn}
\usepackage{bm}

\begin{document}

\preprint{APS/123-QED}

\title{ Revisiting $^{63}$Cu NMR evidence for charge order in superconducting La$_{1.885}$Sr$_{0.115}$CuO$_4$}


 \author{T.\ Imai} 
\affiliation{Department of Physics and Astronomy, McMaster University, Hamilton, Ontario L8S4M1, Canada}
\affiliation{Canadian Institute for Advanced Research, Toronto, Ontario M5G1Z8, Canada}
\author{S.\ K.\ Takahashi} 
\affiliation{Department of Physics and Astronomy, McMaster University, Hamilton, Ontario L8S4M1, Canada}
\author{A.\ Arsenault} 
\affiliation{Department of Physics and Astronomy, McMaster University, Hamilton, Ontario L8S4M1, Canada}
\author{A.\ W.\ Acton} 
\affiliation{Department of Physics and Astronomy, McMaster University, Hamilton, Ontario L8S4M1, Canada}
\author{D.\ Lee} 
\affiliation{Department of Physics and Astronomy, McMaster University, Hamilton, Ontario L8S4M1, Canada}
\author{W.\ He} 
\affiliation{Stanford Institute for Materials and Energy Sciences, Stanford National Accelerator Laboratory, Menlo Park, CA 94025}
\affiliation{Department of Applied Physics, Stanford University, Stanford, CA 94305}
\author{Y.\ S.\ Lee} 
\affiliation{Stanford Institute for Materials and Energy Sciences, Stanford National Accelerator Laboratory, Menlo Park, CA 94025}
\affiliation{Department of Applied Physics, Stanford University, Stanford, CA 94305}
\author{M.\ Fujita} 
\affiliation{Institute for Materials Research, Tohoku University, Sendai, Japan}

\date{\today}

\begin{abstract}
The presence of charge and spin stripe order in the La$_2$CuO$_4$-based family of superconductors continues to lead to new insight on the unusual ground state properties of high $T_c$ cuprates.  Soon after the discovery of charge stripe order at $T_{charge} \simeq 65$~K in Nd$^{3+}$ co-doped La$_{1.48}$Nd$_{0.4}$Sr$_{0.12}$CuO$_4$ ($T_{c}\simeq6$~K) [Tranquada et al., Nature {\bf 375} (1995) 561], Hunt, Singer et al. demonstrated that La$_{1.48}$Nd$_{0.4}$Sr$_{0.12}$CuO$_4$ and superconducting La$_{2-x}$Sr$_{x}$CuO$_4$~with $x \sim 1/8$ ($T_{c}\simeq30$~K) share nearly identical NMR anomalies near $T_{charge}$ of the former [Phys. Rev. Lett. {\bf 82} (1999) 4300].  Their inevitable conclusion that La$_{1.885}$Sr$_{0.115}$CuO$_4$ also undergoes charge order at a comparable temperature became controversial, because diffraction measurements  at the time were unable to detect Bragg peaks associated with charge order.  Recent advances in x-ray diffraction techniques finally led to definitive confirmations of the charge order Bragg peaks in La$_{1.885}$Sr$_{0.115}$CuO$_4$ with an onset at as high as $T_{charge} \simeq 80$~K.  Meanwhile, improved instrumental technology has enabled routine NMR measurements that were not feasible two decades ago.  Motivated by these new developments, we revisit the charge order transition of a La$_{1.885}$Sr$_{0.115}$CuO$_4$ single crystal based on $^{63}$Cu NMR techniques.  We  demonstrate that $^{63}$Cu NMR properties of the nuclear spin $I_{z} = -\frac{1}{2}$ to $+\frac{1}{2}$ central transition below $T_{charge}$ exhibit unprecedentedly strong dependence on the measurement time scale set by the separation time $\tau$ between the 90$^{\circ}$ and 180$^{\circ}$ radio frequency pulses; a new kind of anomalous, very broad wing-like $^{63}$Cu NMR signals gradually emerge below $T_{charge}$ only for extremely short $\tau \lesssim  4~\mu$s, while the spectral weight $I_{Normal}$ of the normal NMR signals is progressively wiped out.  The NMR linewidth and relaxation rates depend strongly on $\tau$ below $T_{charge}$, and their enhancement in the charge ordered state indicates that charge order turns on strong but inhomogeneous growth of Cu spin-spin correlations.  

\end{abstract}

\maketitle


\section{\label{sec:level1} Introduction}
The mysterious properties of the hole-doped CuO$_2$ planes in copper-oxide high $T_c$ superconductors continue to pose a major intellectual challenge three decades after their initial discovery.  The non-trivial nature of the interplay between the charge and spin degrees of freedom in cuprates was vividly displayed by the early discovery of the so-called 1/8 anomaly \cite{Axe, Crawford}.  The superconducting transition temperature $T_c$ in La$_{2-x-y}$Nd$_{y}$Sr$_{x}$CuO$_4$ as well as La$_{2-x}$Ba$_{x}$CuO$_4$ is strongly suppressed for the doping concentration near $x \sim1/8$, and these materials enter into an incommensurate spin density wave (I-SDW) ordered phase \cite{Luke, Nachumi}.  

In 1995, Tranquada et al. used neutron scattering techniques to demonstrate that La$_{1.48}$Nd$_{0.4}$Sr$_{0.12}$CuO$_4$ undergoes successive phase transitions into a charge and spin ordered {\it stripe phase} \cite{Tranquada, TranquadaPRB54, TranquadaPRB59}: the low temperature orthorhombic (LTO) to low temperature tetragonal (LTT) structural phase transition at $T_{LTT} \sim 70$~K is followed by charge order at $T_{charge} \sim 65$~K, then an I-SDW order at $T_{spin}^{neutron} \sim 50$~K.   Due to the glassy nature of the spin order, however, Cu spins continue to fluctuate slowly below $T_{spin}^{neutron}$ and $\mu$SR techniques detect the static spin order only below $T_{spin}^{\mu SR} \sim 35$~K \cite{Nachumi}.  This is because the measurement time scale of $\mu$SR ($\sim 10^{-7}$~s) is slower than that of elastic neutron scattering ($\sim 10^{-11}$~s).  

The charge order in cuprates is rather subtle, and proved to be elusive.  For example, if one conducts $^{63}$Cu NMR measurements on La$_{1.48}$Nd$_{0.4}$Sr$_{0.12}$CuO$_4$ and related materials with the pulse separation time $\tau = 10~\mu$s or greater between the 90$^{\circ}$ excitation and 180$^{\circ}$ refocusing radio frequency pulses, as was usually the case in the late 1980's or 1990's, one can easily overlook any hint of charge order \cite{Imai1990}.  In 1999, Hunt, Singer and co-workers identified NMR anomalies at charge order transition in La$_{1.48}$Nd$_{0.4}$Sr$_{0.12}$CuO$_4$ \cite{HuntPRL, SingerPRB, HuntPRB, Cederstrom, SingerLT23}.  The most striking feature was that $^{63}$Cu NMR signal intensity measured in zero magnetic field using the nuclear quadrupole resonance (NQR) techniques with $\tau \gtrsim 10~\mu$s is gradually wiped out below $T_{charge}$, as reproduced in Fig.\ 1(a) \cite{HuntPRL, SingerPRB}.  The partial disappearance of the $^{63}$Cu NMR signal implies that,  in some segments of the CuO$_2$ planes, the relaxation times of the $^{63}$Cu nuclear spins become too fast and/or their resonant frequency shifts outside the observation window \cite{HuntPRL}, but the details of the mechanism behind the intensity anomaly remained unknown because one cannot characterize unobservable signals.  

Hunt et al. also observed analogous $^{63}$Cu NMR intensity wipeout in La$_{1.88}$Ba$_{0.12}$CuO$_4$ and La$_{1.68}$Eu$_{0.2}$Sr$_{0.12}$CuO$_4$ in the LTT phase, as reproduced in Fig.\ 1(b-c).  These intensity anomalies are accompanied by the enhancement of the low frequency Cu spin fluctuations averaged over the entire volume of the sample, as reflected on the $^{139}$La nuclear spin-lattice relaxation rate $1/T_1$ \cite{HuntPRB, ImaiAspen}.  Moreover, the transverse spin echo decay loses the Gaussian component and becomes Lorentzian (i.e. exponential) for longer values of $\tau$ \cite{HuntPRL, SingerPRB, HuntPRB}.

Interestingly, Hunt et al. found nearly identical NMR anomalies even in the superconducting La$_{1.885}$Sr$_{0.115}$CuO$_4$ \cite{HuntPRL, ImaiAspen}, although it does not undergo the LTO-LTT structural phase transition.  The commonalities of the NMR anomalies inevitably led us to conclude that La$_{1.88}$Ba$_{0.12}$CuO$_4$, La$_{1.68}$Eu$_{0.2}$Sr$_{0.12}$CuO$_4$, and even the superconducting La$_{1.885}$Sr$_{0.115}$CuO$_4$ undergo a charge order transition with comparable $T_{charge}$ as La$_{1.48}$Nd$_{0.4}$Sr$_{0.12}$CuO$_4$.  Our unexpected findings surprised the high $T_c$ community \cite{ServiceScience}.  

Subsequently, Abu-Shiekah et al. also reported very similar  $^{139}$La NMR anomalies in La$_2$NiO$_{4.17}$ below the charge order temperature that was independently determined by Bragg scattering \cite{Abu-Shiekah1999}.  ($^{139}$La NMR could probe the charge and spin order in NiO$_2$ planes quite effectively, owing to strong hybridization between the Ni$^{2+}$ 3d$_{3z^{2}-r^{2}}$ orbital and La sites.  But $^{139}$La NMR is less effective in cuprates, because Cu$^{2+}$ 3d$_{x^{2}-y^{2}}$ orbital extends only within the CuO$_2$ planes.)  Despite the clear links established between the NMR anomalies and charge order known for La$_{1.48}$Nd$_{0.4}$Sr$_{0.12}$CuO$_4$ and La$_2$NiO$_{4.17}$, our conclusion for the presence of charge order in La$_{1.885}$Sr$_{0.115}$CuO$_4$, La$_{1.88}$Ba$_{0.12}$CuO$_4$ and La$_{1.68}$Eu$_{0.2}$Sr$_{0.12}$CuO$_4$ later became controversial, because the world-wide effort to search for the Bragg scattering signals associated with charge order failed at the time in these three materials.       

\begin{figure}
\centering
\includegraphics[width=3.5in]{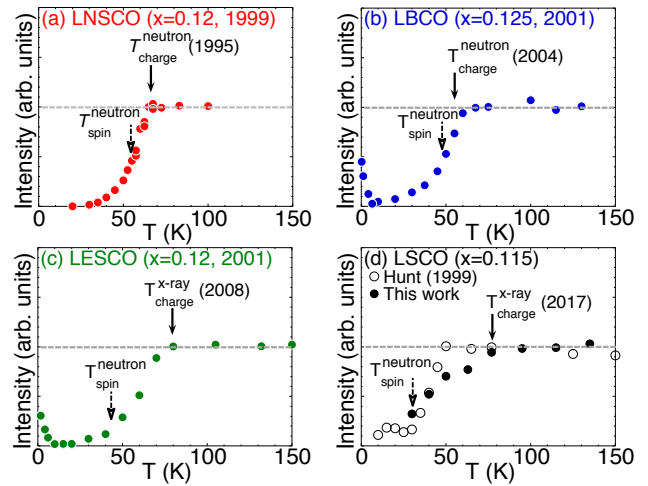}
\caption{(a) The temperature dependence of the integrated intensity of $^{63}$Cu NMR lineshapes in La$_{1.48}$Nd$_{0.4}$Sr$_{0.12}$CuO$_4$, measured in zero magnetic field using the nuclear quadrupole resonance (NQR) techniques \cite{HuntPRL, SingerPRB}.  The intensity is corrected for the Boltzmann factor, and for the transverse spin echo decay measured for $2\tau > 24 ~\mu$s.  The arrow marks the onset of charge order at $T_{charge}^{neutron} \sim 65$~K, whereas the dashed arrow marks the spin order at the fast measurement time scale of elastic neutron scattering, $T_{spin}^{neutron} \sim 55$~K,  both as determined by Tranquada et al. \cite{Tranquada, TranquadaPRB54, TranquadaPRB59}.   (b - c) The NQR results for La$_{1.88}$Ba$_{0.12}$CuO$_4$ \cite{HuntPRL, HuntPRB} and La$_{1.68}$Eu$_{0.2}$Sr$_{0.12}$CuO$_4$ \cite{HuntPRL, HuntPRB}, compared with $T_{charge}^{neutron, x-ray}$ subsequently determined by neutron and x-ray scattering technique, respectively \cite{Fujita, Fink, Fink2}.  The signal intensity begins to recover toward the base temperature when the hyperfine magnetic fields arising from the ordered spins become static far below $T_{spin}$ (this feature is missing in (a) due to the influence of Nd$^{3+}$ spin order).  (d)  Open circles ($\circ$): the $^{63}$Cu NQR intensity of La$_{1.885}$Sr$_{0.115}$CuO$_4$ reported in 1999 \cite{HuntPRL}, which underestimated  the onset of charge order \cite{He}.  Filled circles ($\bullet$): the new single crystal NMR results of the spectral weight of the normal $^{63}$Cu NMR central peak, $I_{Normal}$, reported in this work (from Fig.\ 7).   The structural transition to the LTT phase (not shown in the panels) takes place at $T_{LTT} =70$~K, 54~K, and 135~K in (a) La$_{1.48}$Nd$_{0.4}$Sr$_{0.12}$CuO$_4$, (b) La$_{1.88}$Ba$_{0.12}$CuO$_4$ and (c) La$_{1.68}$Eu$_{0.2}$Sr$_{0.12}$CuO$_4$, respectively.  
}\label{cs}
\end{figure}

Fujita et al. were the first to detect the charge order Bragg peaks of La$_{1.88}$Ba$_{0.12}$CuO$_4$ successfully below $T_{charge} \sim 54$~K based on neutron diffraction measurements \cite{Fujita}.  More recently, a new generation of x-ray scattering experiments finally led to  confirmation of charge order also in La$_{1.8-x}$Eu$_{0.2}$Sr$_{x}$CuO$_4$ ($T_{charge}^{x-ray} \simeq 80$~K) \cite{Fink} and superconducting La$_{1.885}$Sr$_{0.115}$CuO$_4$ ($T_{charge}^{x-ray} \simeq 80$~K) \cite{Thampy, Croft, He}.  In Fig.\ 1, we compare $T_{charge}$ as determined by diffraction measurements with the temperature dependence of the $^{63}$Cu NMR signal intensity wipeout we reported two decades ago.  The agreement is very good for La$_{1.48}$Nd$_{0.4}$Sr$_{0.12}$CuO$_4$, La$_{1.88}$Ba$_{0.12}$CuO$_4$, and La$_{1.68}$Eu$_{0.2}$Sr$_{0.12}$CuO$_4$.  It turned out, however, that charge order in La$_{1.885}$Sr$_{0.115}$CuO$_4$ sets in gradually, starting at as high as $T_{charge}^{x-ray} \simeq 80$~K \cite{Croft, Thampy, He}, whereas the original powder NQR data by Hunt et al. suggested a much sharper charge order transition at $T_{charge} \simeq 50$~K \cite{HuntPRL}.  Clearly, Hunt et al. overlooked the gradual onset of the charge order transition.  

These new developments motivated us to revisit NMR signatures of charge order in superconducting La$_{1.885}$Sr$_{0.115}$CuO$_4$ ($T_{c} = 30$~K) based on $^{63}$Cu NMR techniques, using a single crystal with a known charge order temperature $T_{charge}\simeq 80$~K that we determined independently by x-ray scattering experiments \cite{He}.  As explained in detail in section II, advances in digital electronics technologies have enabled routine NMR measurements possible with extremely short $\tau = 2~\mu$s, effortlessly.  In what follows, we will demonstrate that the main NMR peak begins to lose the spectral weight $I_{Normal}$ precisely below $T_{charge}\simeq 80$~K, because very broad, anomalous wing-like NMR signals gradually emerge in the charge ordered state.  The wing-like signals are observable only when we conduct NMR measurements with a very fast ``shutter speed" set by $\tau \simeq 2~\mu$s.  From the measurements of the NMR linewidths and relaxation rates, we will show that Cu spin-spin correlations are enhanced strongly but inhomogeneously below $T_{charge}$ in a growing volume fraction of the CuO$_2$ planes. 

\section{\label{sec:level1} Experimental }

We grew a single crystal of La$_{1.885}$Sr$_{0.115}$CuO$_4$ with traveling solvent floating zone techniques at Tohoku.  We aligned the crystal with Laue techniques at Stanford, and cut it to a rectangular shape with the approximate dimensions of 2.5~mm $\times$ 2.5~mm $\times$ 1mm.  Susceptibility measurements using a superconducting quantum interference device (SQUID) showed a sharp bulk superconducting transition at $T_{c}=30$~K, which is known to coincide with the onset of spin order at the time scale of elastic neutron scattering, $T_{spin}^{neutron}\simeq30$~K \cite{Kimura}.  An analogous crystal cut from the same boule was used for high precision x-ray diffraction experiments at the SLAC, and exhibited a gradual onset of charge order below $T_{charge}\simeq 80$~K \cite{He}.  

We conducted all the NMR measurements at McMaster using a state-of-the-art NMR spectrometer built around the Redstone NMR console acquired from Tecmag Inc.  The metallic single crystal inside the NMR coil strongly damps the Q-factor of the tank circuit, and hence the tail end of the high voltage radio-frequency pulses decays in less than $1~\mu$s after we turn the radio frequency pulses off, as monitored in-situ using a pick-up antenna.  The inherently low Q was useful in suppressing the saturation of the preamplifier for signal detection.   We applied the radio-frequency pulses ($H_{1}$) along the ab-plane of the crystal.  This ensures that the joule current would have to loop along the c-axis as well as within the ab-plane, and the resistivity along the c-axis is three orders of magnitude larger (and increases with decreasing temperature) \cite{SuzukiResistivity}.  This geometry is known to work well for the NMR intensity measurements \cite{TakedaPRL}.  In fact, within the experimental uncertainties, the integrated intensity of the NMR lineshape measured with the extremely short delay time $\tau = 2~\mu$s is conserved except near $T_{spin}^{neutron}$ (Fig.\ 7(a) below), although the in-plane resistivity $\rho_{ab}$ decreases with temperature \cite{SuzukiResistivity}.

We used the model LN-2M acquired from Doty Inc. as the preamplifier for spin echo detection.  The recovery time of the preamplifier after high voltage saturation is $\sim 1~\mu$s.  The duration of the overall spectrometer dead time, $t_{dead}$, caused by the ring down from the saturated preamplifier is as short as $t_{dead}\lesssim 2.5~\mu$s after we turn the 180$^{\circ}$ pulse off.  

Such a short $t_{dead}$ of our modern NMR spectrometer is a major advantage over the previous generation of NMR spectrometers built in the 1990's around the Aries console (also from Tecmag inc.); the typical spectrometer dead time was $t_{dead} = 8 \sim 12~\mu$s for high field NMR measurements at $\sim 100$~MHz and $t_{dead} = 10 \sim 12~\mu$s for NQR measurements at lower frequencies ($30\sim 40$~MHz) for our Aries based spectrometers.  In the 1990's, the longer $t_{dead}$ prevented us from detecting $^{63}$Cu NMR signals arising from the charge ordered segments of the CuO$_2$ planes below $T_{charge}$.  This is because when the NMR relaxation times become shorter than $t_{dead}$, we lose the NMR signals, as schematically explained in Fig.\ 2.  A major thrust of the present work is that we successfully detected and characterized a new kind of anomalous $^{63}$Cu NMR signals that emerges below $T_{charge}$ only for $\tau \lesssim 4~\mu$s.

\begin{figure}
\centering
\includegraphics[width=3in]{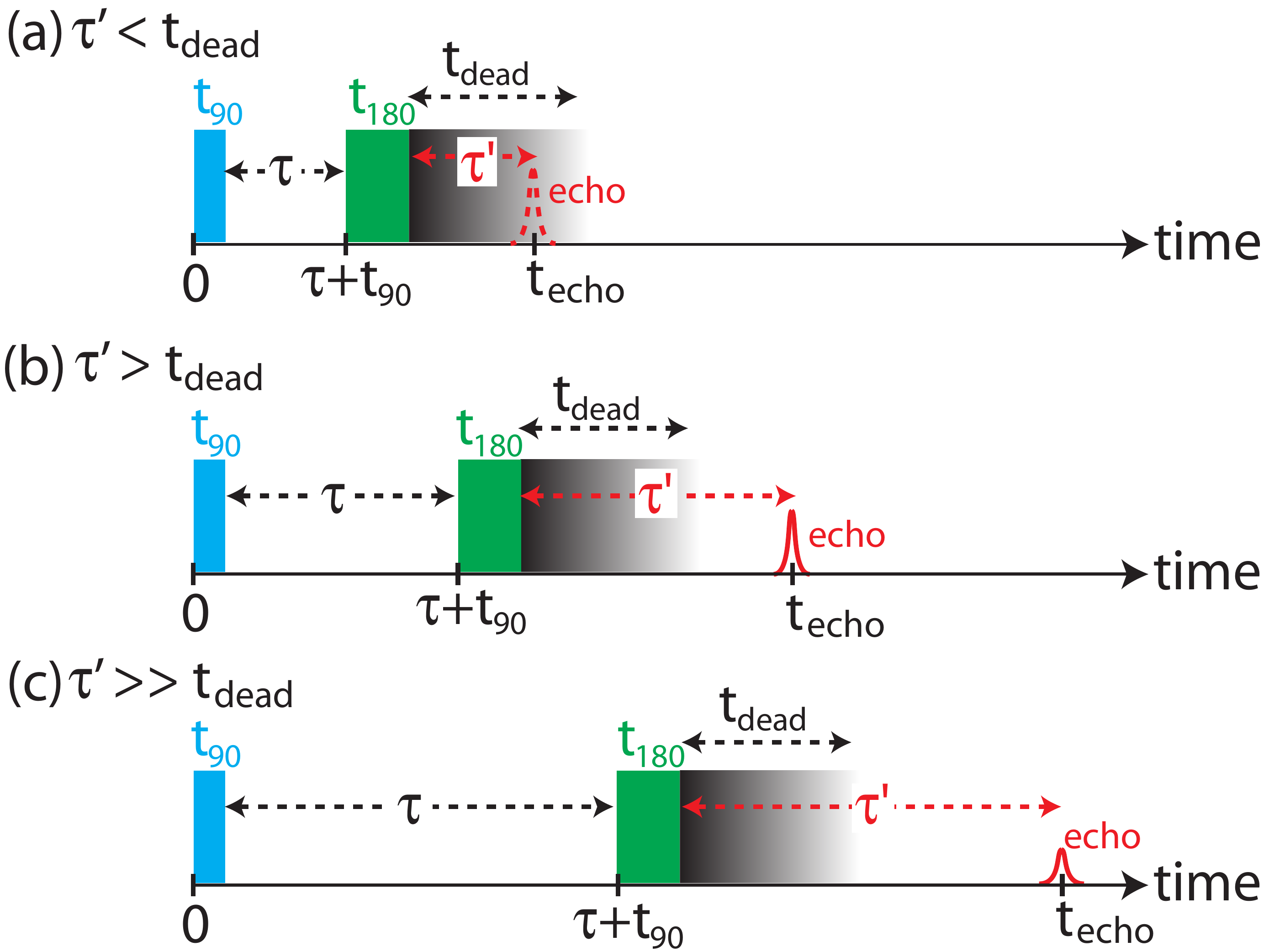}
\caption{A schematic representation of the spin echo measurement using the 90$^{\circ}$ excitation and 180$^{\circ}$ refocusing radio-frequency pulses separated by a delay time $\tau$.  The typical duration of these pulses in this work is $t_{90} = 2.5~\mu$s and $t_{180} = 5~\mu$s, respectively.    A spin echo appears at $\tau'=\tau + t_{180}/\pi$ after we turn the 180$^{\circ}$ pulse off \cite{Slichter}.   (a) If $\tau'$ is shorter than the spectrometer dead time $t_{dead}$, we are unable to observe the spin echo signal.  (b) If $\tau' > t_{dead}$, we can observe a spin echo signal after the dead time.  (c) Even if $\tau' > t_{dead}$, the spin echo intensity is suppressed when the transverse relaxation time $T_2 \ll \tau$.
}\label{cs}
\end{figure}

\section{\label{sec:level1} Results and Discussions}
\subsubsection{$^{63}{Cu}$ NMR spin echo signals}

\begin{figure}
\centering
\includegraphics[width=3.2in]{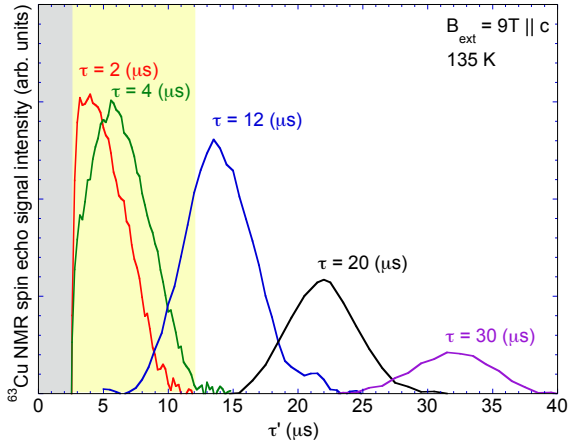}
\caption{Representative time traces of the spin echo signal observed at 135~K for the delay time $\tau = 2$ to $30~\mu$s at the peak of the $^{63}$Cu NMR lineshape, $f_{o} = 102.74$~MHz.  The pulse width is $t_{180} = 5~\mu$s in these measurements, and hence the maximum of the spin echo appears at $\tau'\sim (\tau + 1.7)~ \mu$s for a given $\tau$.  The region marked with gray shade represents  the time domain inaccessible in our NMR experiments due to the spectrometer dead time, $t_{dead} \sim 2.5~\mu$s, whereas the yellow shade marks the longer dead time $t_{dead} \sim 12~\mu$s encountered in our earlier $^{63}$Cu NQR experiments in the late 1990's \cite{HuntPRL}.  
}\label{cs}
\end{figure}

In Fig.\ 3, we present the typical time traces of the $^{63}$Cu NMR spin echo signal observed at 135~K for various delay times between $\tau = 2~\mu$s and 30~$\mu$s.  As explained in Fig.\ 2, the peak of the spin echo signal appears at $\tau' =\tau + t_{180}/\pi$ after we turn the 180$^{\circ}$ pulse off.  In order to detect the peak of the spin echo signal in the time domain properly without suffering from the non-linearity during the spectrometer dead time (marked with grey shade), we need to maintain $\tau' > t_{dead}$.  Since $t_{180} \simeq 5~\mu$s and $t_{dead} \gtrsim 2.5~\mu$s, we were able to reduce $\tau$ to $2~\mu$s.   

\begin{figure}
\centering
\includegraphics[width=3.2in]{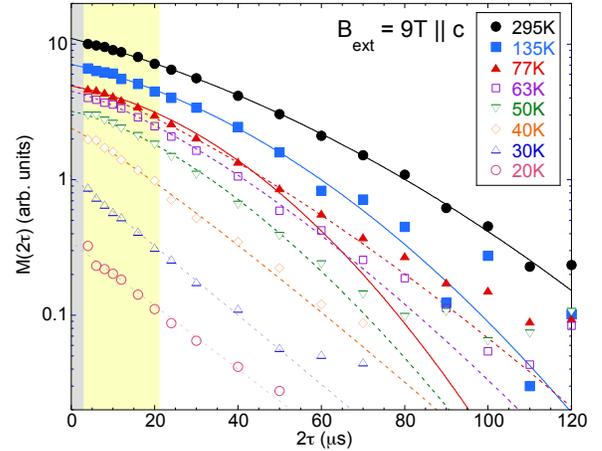}
\caption{Examples of $^{63}$Cu spin echo decay curves $M(2\tau)$ measured at the center of the NMR lineshape with an external magnetic field $B_{ext} = 9$~T applied along the c-axis.  All the signal intensities are corrected for the Boltzmann factor by multiplying temperature $T$.  Solid curves are the Lorentzian-Gaussian fit with Eq.~(6) with the fixed $T_{2R}$ estimated from $T_1$ measurements.  As we approach $T_{charge}$ ($\simeq 80$~K) from higher temperatures, the fit becomes poor for longer values of $2\tau$ due to the disappearance of the Gaussian curvature.  Dashed curves are guides for eyes based on the free Lorentzian-Gaussian fit without the constraint on $T_{2R}$.
}\label{cs}
\end{figure}

We determined $M(2\tau)$, the magnitude of the spin echo signal at its peak for a given $\tau$, by integrating the spin echo signal around its peak in the time domain.  In Fig.\ 4, we summarize the spin echo decay $M(2\tau)$ as a function of $2\tau$ for representative temperatures.  We normalized the overall magnitude of $M(2\tau)$ by multiplying temperature $T$, to take into account the effect of the Boltzmann factor on the signal intensities.  The extrapolation of $M(2\tau)$ to $\tau = 0$, $M(0)$, is proportional to the number of nuclear spins detected from the sample.  $M(0)$ appears to decrease even below 295~K down to $T_{charge} \simeq 80$~K, simply because the width of the NMR lineshape broadens in the frequency domain.

\subsubsection{$^{63}{Cu}$ NMR lineshapes}
In Fig.\ 5(a), we show the $^{63}$Cu NMR lineshapes of the nuclear spin $I_{z} = +1/2$ to $-1/2$ central transition observed at 135~K ($ > T_{charge}$) using various values of $\tau$ in an external magnetic field of $B_{ext}=9.0$~T applied along the crystal c-axis.  In general, the resonant frequency $\protect{f_{o}}$ of the central transition may be written as
\begin{equation}
          f_{o} = \gamma_{n} (1+K^{(c)})B_{ext} + \Delta \nu_{Q}^{(2)},
\label{2}
\end{equation}
where $\gamma_{n}/2\pi = 11.285$~MHz/T is the nuclear gyromagnetic ratio of the $^{63}$Cu nuclear spin.  $\Delta \nu_{Q}^{(2)}$ arises from the second order effect of the nuclear quadrupole interaction, which is inversely proportional to $B_{ext}$.  Since the main principal axis of the electric field gradient (EFG) tensor for La$_{1.885}$Sr$_{0.115}$CuO$_4$ is parallel with the c-axis and the asymmetry of the EFG tensor is negligibly small,  $\Delta \nu_{Q}^{(2)} \simeq 0$ for the $B_{ext}~||~c$-axis geometry.  

\begin{figure}
\centering
\includegraphics[width=3.5in]{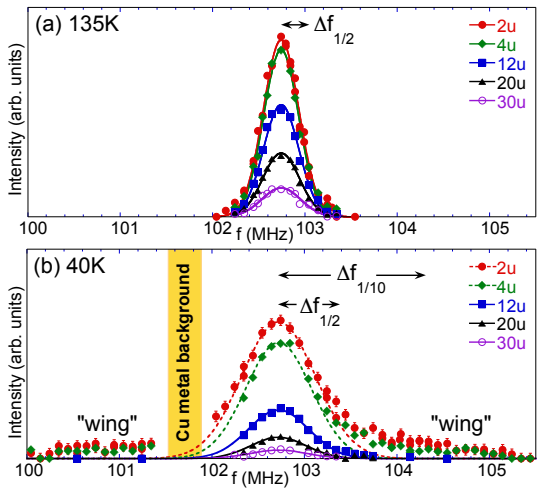}
\caption{(a) $^{63}$Cu NMR lineshapes of the $I_{z} = +1/2$ to $-1/2$ central transition measured at 135~K in a magnetic field $B_{ext} = 9$~T applied along the c-axis for various delay times $\tau = 2~\mu$s to 30~$\mu$s.  Solid curves are the best Gaussian fit of the entire lineshape with the half width at the half maximum (HWHM) $\Delta f_{1/2} = 278 \pm 18$~kHz.  The lineshapes are independent of $\tau$, except for the overall magnitude.  (b) At 40~K, wing-like NMR signals emerge at around 100.5~MHz and 104.5~MHz for very short $\tau$.  Notice that the Gaussian fit of the narrower main peak around 102.7~MHz underestimates the wing-like signals for $\tau = 2~\mu$s and 4~$\mu$s, as shown by dashed curves.  For longer $\tau$, the wing-like signals disappear, and the Gaussian lineshape is recovered.  The light orange band near 101.8~MHz represents a frequency range, where background signals from $^{63}$Cu metal in the probe and the resonant coil made accurate measurements difficult.  $\Delta f_{1/2}$ and  $\Delta f_{1/10}$ are the half width at the 1/2 intensity and at the 1/10 intensity, respectively.
}\label{cs}
\end{figure}

\begin{figure}
\includegraphics[width=3.4in]{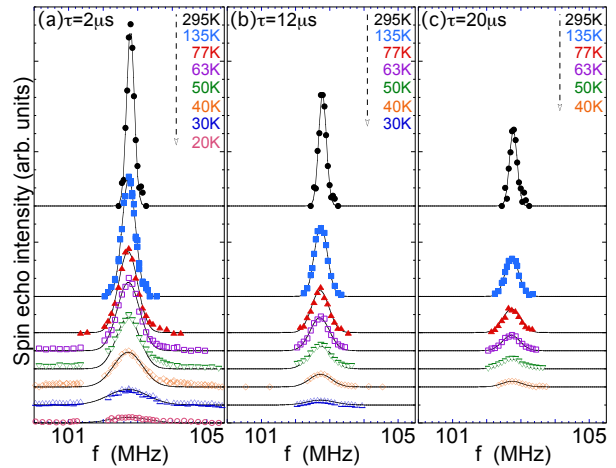}
\caption{(a) $^{63}$Cu NMR lineshapes measured with $\tau = 2 ~\mu$s at various temperatures.  All the lineshapes are normalized for the Boltzman factor by multiplying temperature $T$.  For clarity, we shifted the origin vertically at different temperatures.  Notice that wing-like signals appear below $T_{charge}$, and their frequency range extends with decreasing temperature.  The Gaussian fit of the narrower main peak (solid line) underestimates the intensity of wings.  (b - c) Aside from the mild broadening, the only anomaly observed below $T_{charge}$ for longer values of  $\tau = 12~\mu$s and $\tau = 20~\mu$s is the loss of the signal intensity.  Gaussian fits work very well in the absence of wing-like signals even below $T_{charge}$.
}\label{cs}
\end{figure}

\begin{figure}
\centering
\includegraphics[width=3.5in]{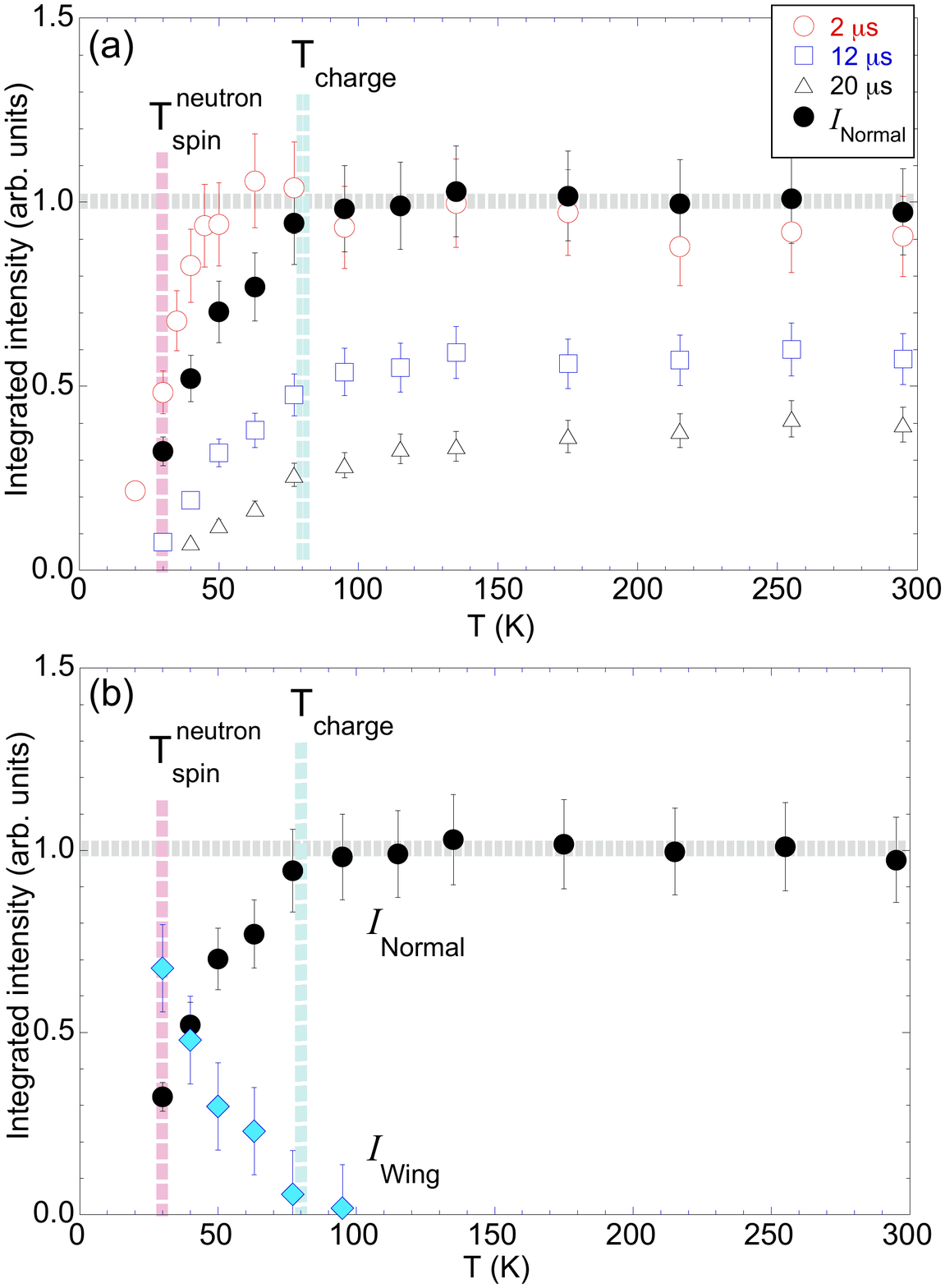}
\caption{(a) Open symbols: the integrated intensity of the $^{63}$Cu NMR lineshapes for fixed $\tau = 2$, 12, and  20 $\mu$s in Fig.\ 6.  Filled circles: $I_{Normal}$, the spectral weight of the normally behaving narrower main peak around $\protect{f_{o}} \simeq 102.7$~MHz, as estimated by extrapolating the integrated intensity observed at $\tau = 12~\mu$s to $\tau=0$ using the spin echo decay curves summarized in Fig.\ 4.  (b) The spectral weight of the anomalous wing-like segments, $I_{Wing} = 1-I_{Normal}$, in comparison to $I_{Normal}$.
}\label{cs}
\end{figure}

$K^{(c)}$ in eq.(1) is the NMR frequency shift (also known as the Knight shift), and may be divided into the spin contribution $K_{spin}^{(c)}$ and the temperature independent orbital contribution $K_{orb}^{(c)}$ as  
\begin{eqnarray}
K^{(c)} = K_{spin}^{(c)} + K_{orb}^{(c)},
\end{eqnarray}
\begin{eqnarray}
K_{spin}^{(c)} = \frac{A_{hf}^{(c)}({\bf q}={\bf 0})}{N_{A}\mu_{B}}\chi_{spin}^{(c)},
\end{eqnarray}
where $N_{A}$ is Avogadro's number,  $A_{hf}^{(c)}({\bf q}={\bf 0})$ is the wave vector ${\bf q}={\bf 0}$ component of the  form factor for the hyperfine interactions, and $\chi_{spin}^{(c)}$ represents the local spin susceptibility.  

In YBa$_{2}$Cu$_{3}$O$_7$ \cite{Mila}, it is well known that the negative contribution of the on-site hyperfine interaction $A_{c} \simeq -16$~(T/$\mu_{B}$) accidentally cancels out with the positive contributions from the super-transferred hyperfine interaction $B \simeq 4$~(T/$\mu_{B}$) with four neighboring Cu sites, $A_{hf}^{(c)}({\bf q}={\bf 0}) = A_{c}+4B \simeq 0$.  Accordingly, $K_{spin}^{(c)} \simeq 0$ and the peak frequency of the central transition in the c-axis geometry is set almost entirely by $K_{orb}^{(c)} \simeq 1.28$~\% \cite{TakigawaPRB1989shift, Barrett}.  An analogous situation is realized also for the paramagnetic state of La$_2$CuO$_{4}$ and the Sr$^{2+}$ doped variants \cite{Imai1990, Imai1993_1, Imai1993_2}.  This is why the $^{63}$Cu NMR peak frequency hardly changes from $f_{o} \simeq 102.7$~MHz between 135~K and 40~K in   Fig.\ 5 despite a significant decrease of $\chi_{spin}^{(c)}$ with temperature \cite{Johnston}.

The peak intensity in Fig.\ 5(a) becomes smaller for longer values of $\tau$ due to the transverse $T_2$ relaxation process, as summarized in Fig.\ 4; otherwise, the lineshapes remain identical for different values of $\tau$ above $T_{charge}$.  We can fit the entire lineshape nicely with a Gaussian function with a constant half width at the half maximum (HWHM), $\Delta f_{1/2} = 278 \pm 18$~kHz regardless of $\tau$.  These findings above $T_{charge}$ are quite normal.  

In contrast, the lineshapes in Fig.\ 5(b) measured at 40~K in the charge ordered state show unprecedentedly strong dependence on $\tau$.  The HWHM of the main peak becomes larger for shorter values of $\tau$.   In addition, wing-like symmetrical NMR signals emerge on both higher and lower frequency sides of the main peak below $\tau \sim 4 \mu$s.  We found that the spin echo decay of the wing-like signals is pure Lorentzian (i.e. exponential) and the transverse relaxation time is as fast as $T_{2} = 10.5~\mu$s at 104.14~MHz; this $T_{2}$ is shorter than the typical $t_{dead}$ during the 1990's, and hence everyone overlooked the $^{63}$Cu NMR signals arising from nuclear spins belonging to the wing-like segments. 

We summarize the temperature dependence of the NMR lineshapes for fixed $\tau = 2~\mu$s in Fig.\ 6(a).   We can see evolution of the wing-like signals below $T_{charge}$.  The integrated intensity of the entire lineshape is conserved from 295~K down to $\sim 40$~K through $T_{charge}$, as shown  in Fig.\ 7(a).  This means that the lost spectral weight from the narrower main peak around $\protect{f_{o}} \simeq 102.7$~MHz is transferred to the wing-like segments below $T_{charge}$.  In the case of longer $\tau = 12~\mu$s, and $20~\mu$s in Fig.\ 6 (b-c), the wing-like segments are missing due to the short $T_{2}$ and the integrated intensity drops quickly below $T_{charge}$.

\subsubsection{$^{63}{Cu}$ spin-lattice relaxation rate $1/T_1$}        
We measured the nuclear spin-lattice relaxation rate $1/T_1$ using the inversion recovery technique at the peak of the central transition with various values of $\tau$.  See Appendix A for the representative examples of the recovery curves observed at 40~K.  We summarize the $1/T_1$ results in Fig.\ 8(a).    We also compare $1/T_1$ measured at 104.13~MHz for the wing-like NMR signals below $T_{charge}$.  We confirmed that $1/T_1$ measured at somewhat different frequencies within the wing-like segment is not significantly different.    The data points within the area shaded by light blue are measured for residual paramagnetic NMR signals that begin to diminish near and below $T_{spin}^{neutron}$, and hence may represent only a small volume fraction of the CuO$_2$ planes; accordingly, these data points need to be interpreted with caution.  

The $1/T_1$ results measured with longer $\tau = 12$ to $20~\mu$s are qualitatively similar to those observed for the optimally superconducting La$_{2-x}$Sr$_{x}$CuO$_4$ ($x \sim 0.15$) \cite{Ishida, Imai1990, Ohsugi1994} and YBa$_{2}$Cu$_{3}$O$_7$ \cite{Imai1988, Barrett}.   To underscore this point, we plot $T_1$ multiplied by $T$ in Fig.\ 8(b).  In general, $1/T_{1}T$ probes the wave vector $\bf{q}$-integral of the imaginary part of the dynamical electron spin susceptibility, $\chi''(\bf{q}, f_{o})$, at the NMR frequency $f_{o}$\cite{Moriya1963}.  We found that our results for $\tau = 20~\mu$s fits  nicely with a Curie-Weiss form,
\begin{equation}
          1/T_{1}T \sim \frac{C}{T + \theta},
\label{2}
\end{equation} 
with a Weiss temperature $\theta = 130 \pm 15$~K.  Such a Curie-Weiss behavior of $1/T_{1}T$ with positive $\theta$ signals the growth of antiferromagnetic Cu electron spin-spin correlations within the CuO$_2$ plane \cite{MMP1990}.  Analogous Curie-Weiss behavior was previously reported for YBa$_{2}$Cu$_{3}$O$_7$ and La$_{2-x}$Sr$_{x}$CuO$_4$ \cite{MMP1990, Barrett, Ohsugi1994}. 

\begin{figure}
\includegraphics[width=3.5in]{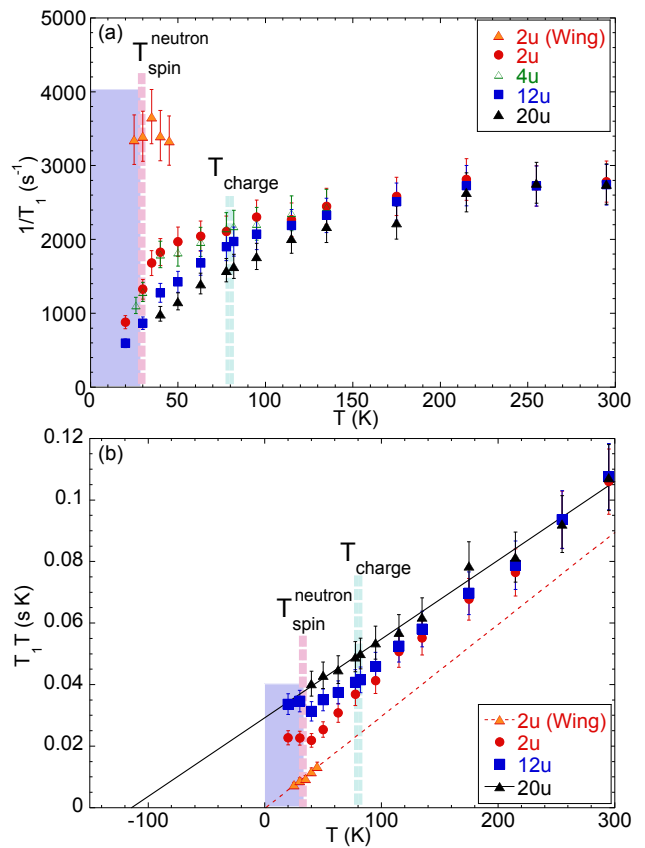}
\caption{(a) The spin-lattice relaxation rate, $1/T_{1}$ and (b) $T_{1}T$, measured at the center of the main peak for various values of $\tau$ between 2~$\mu$s and 20~$\mu$s.  For $\tau = 2~\mu$s, we also show the $1/T_{1}$ results for the wing-like signals below $T_{charge}$ measured at 104.13 MHz.  The black solid line through $T_{1}T$ is the best Curie-Weiss fit with $\theta = 130 \pm 15$~K for $\tau = 20~\mu$s, whereas the red dashed line is a provisional fit for  the wing-like signal with $\theta \sim 0$.  In this and other figures throughout this paper, the region with light-blue shade represents a low temperature range, where we detect only a small fraction of the nuclear spins within the sample due to the diminishing signal intensity.   
}\label{cs}
\end{figure}

The $1/T_1$ results measured with shorter $\tau$ are different.  In particular, $1/T_1$ for the wing-like signals is nearly five times faster, indicating that low frequency antiferromagnetic Cu spin fluctuations are much stronger {\it in some segments of the CuO$_2$ planes where these nuclear spins are located}.  This is consistent with the fact that the inelastic neutron scattering signal intensity measured at 0.3~meV \cite{RomerNeutron}, as well as $1/T_{1}T$ measured at $^{139}$La sites \cite{ImaiAspen, Arsenault}, also begin to grow below $T_{charge}$.  It should be cautioned, however, that inelastic neutron scattering measures the volume integral of the overall response, and their observation does not prove that spin fluctuations are {\it uniformly} slowing down.  In fact, the recovery curve of $1/T_{1}$ observed at the $^{139}$La sites show a clear sign of growing distribution of $1/T_1$ precisely below $T_{charge}$ for this composition \cite{ImaiAspen, Arsenault, Mitrovic}.

Even for the narrower main peak, $1/T_1$ measured with $\tau = 2~\mu$s is significantly faster than with $\tau = 20~\mu$s, but this trend persists even above $T_{charge}$.  Our earlier $^{63}$Cu NQR and $^{17}$O NMR measurements demonstrated that random substitution of Sr$^{2+}$ ions induces quenched disorder, and mild inhomogeneity of local hole concentration $x_{local}$ exists within the CuO$_2$ planes; such a patch by patch distribution of $x_{local}$ has a length scale of the order of several nm \cite{SingerLSCOPRL, SingerPRB2005}.  In other words, the magnitude of $1/T_1$ in La$_{2-x}$Sr$_{x}$CuO$_4$ varies position by position within the CuO$_2$ plane below room temperature even without charge order; the greater $x_{local}$, the slower $T_1$ and $T_2$.   With the current $B_{ext}~||~c$-axis field geometry, all the paramagnetic $^{63}$Cu NMR signals with different $x_{local}$ are centered and superposed at the same $f_{o} \simeq 102.7$~MHz.  This explains why a mild $\tau$ dependence of $1/T_1$ persists even above $T_{charge}$, because $1/T_1$ measured with a longer $\tau$ tends to have greater relative contributions from the regions with larger $x_{local}$.        

It is also noteworthy that, unlike the case of typical second order magnetic phase transitions,  $1/T_{1}$ does not diverge at $T_{spin}^{neutron} \simeq 30$~K when Cu spins begin to statically order at an extremely fast time scale ($\sim 10^{-11}~$s) of the elastic neutron scattering measurements.  Earlier $\mu$SR measurements demonstrated that Cu magnetic moments continue to fluctuate slowly below $T_{spin}^{neutron}$, and begin to order only below $T_{spin}^{\mu SR} = 15 \sim 20$~K at its slower measurement time scales ($\sim 10^{-7}~$s) \cite{Kumagai, Savici}.  This apparent discrepancy is caused by the same glassiness of I-SDW order as mentioned earlier for La$_{1.48}$Nd$_{0.4}$Sr$_{0.12}$CuO$_4$, but the temperature scale is somewhat smaller in the present case without Nd$^{3+}$ co-doping.  Our NMR measurements have an even slower time scale set by $\tau = 2~\mu$s or longer.  The fact that nearly 50\% of paramagnetic NMR signals remain at $T_{spin}^{neutron} = 30$~K for $\tau = 2~\mu$s implies that magnetic order is not imminent in a half of the volume fraction of the CuO$_2$ planes when superconductivity sets in also at $T_{c} = 30$~K.  In fact, $1/T_1$ measured for the residual NMR signals drops below $T_{c} =30$~K without exhibiting a Hebel-Slichter coherence peak expected for conventional s-wave pairing, as previously reported for the case of bulk superconductivity in YBa$_{2}$Cu$_{3}$O$_{6.9}$ \cite{Imai1988} and La$_{1.85}$Sr$_{0.15}$CuO$_4$ \cite{Ishida}.

\subsubsection{$^{63}$Cu NMR linewidth}

We summarize the temperature dependence of the half width at half maximum, $\Delta f_{1/2}$, of the NMR lineshapes in Fig.\ 9(a), and its inverse in Fig.\ 9(b).  $\Delta f_{1/2}$ in La$_{2-x}$Sr$_{x}$CuO$_4$ is nearly an order of magnitude broader than that observed for YBa$_{2}$Cu$_{3}$O$_7$ \cite{TakigawaPRB39, Barrett}, and grows for larger Sr concentration $x$ \cite{Imai1990}.  The exact cause of the line broadening in La$_{2-x}$Sr$_{x}$CuO$_4$ has not been understood very well since the early days of high $T_c$ superconductivity.  We conducted preliminary lineshape measurements at 6~T and confirmed that $\Delta f_{1/2}$ is proportional to the magnetic field; this rules out the possibility that a distribution of $\Delta \nu_{Q}^{(2)}$ ($\propto 1/B_{ext}$) in Eq.\ (1) is the mechanism of the large temperature dependent $\Delta f_{1/2}$.   

In principle, inhomogeneous line broadening caused by the distributions of $K_{spin}^{(c)}$ \cite{SingerPRB2005} and $K_{orb}^{(c)}$  \cite{Haase} might contribute to $\Delta f_{1/2}$ in the present case, too, as previously proposed for $^{17}$O NMR line broadening.  But it seems unlikely that they are the dominant mechanisms behind the $^{63}$Cu NMR line broadening in this field geometry, because the homogeneous linewidth measured with $1/T_{2G}$ shows identical temperature dependence as $\Delta f_{1/2}$ as discussed in the next section.  We also recall that $A_{hf}^{(c)}({\bf q} ={\bf 0}) \simeq 0$, which suppresses a distribution of $K_{spin}^{(c)}$.  Moreover, the frequency shift reaches negative for the low frequency end of the lineshape, but by default $K_{orb}^{(c)}$ is always positive \cite{Mila}.  This means that a distribution of $K_{orb}^{(c)}$ alone cannot account for the observed broadening.

\begin{figure}
\includegraphics[width=3.5in]{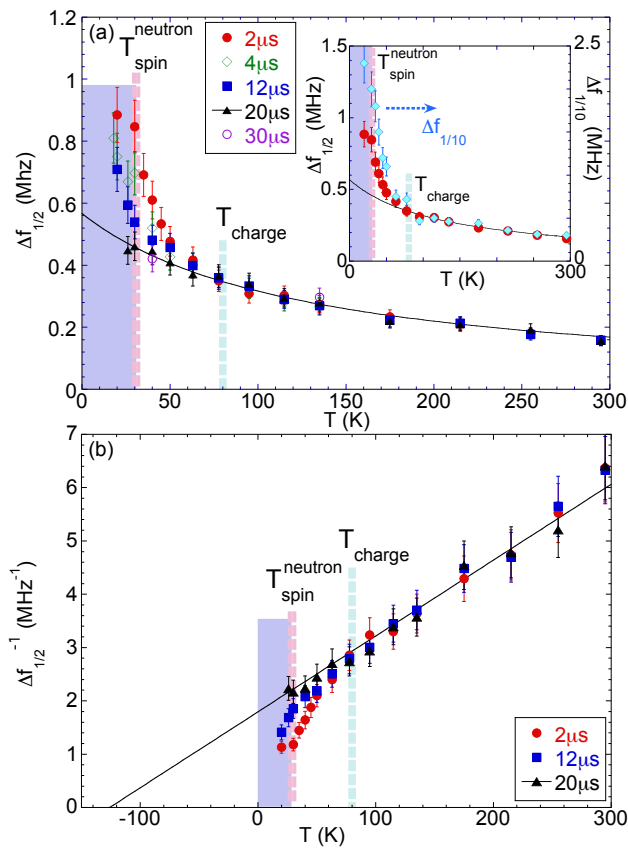}
\caption{(a) Main panel: the half width at the half maximum intensity, $\Delta f_{1/2}$, measured with various values of $\tau$ from $2~\mu$s to $30~\mu$s.  The solid curve is the best Curie-Weiss fit for $\tau = 20~\mu$s with $\theta' = 126 \pm 15$~K.  Inset: comparison of the $\Delta f_{1/2}$ (filled circles) and $\Delta f_{1/10}$ (filled diamonds), both measured for $\tau = 2~\mu$s.  (b) The inverse of $\Delta f_{1/2}$.  Solid line is the same Curie-Weiss fit as in (a).
}\label{cs}
\end{figure}

Above $T_{charge}$, $\Delta f_{1/2}$ does not depend on $\tau$, and the temperature dependence of $\Delta f_{1/2}$ obeys a Curie-Weiss law
\begin{equation}
          \Delta f_{1/2} \sim \frac{C}{T + \theta'}.
\label{2}
\end{equation} 
For the NMR lineshapes measured with longer $\tau$, the wing-like signals are completely suppressed by the very fast transverse relaxation time $T_2$ and the Curie-Weiss law extends below $T_{charge}$.   From the best fit, we found the Weiss temperature $\theta' = 126 \pm 15$~K; this value agrees well with $\theta = 130 \pm 15$~K as determined for the imaginary part of the dynamical electron spin susceptibility as probed by by $1/T_{1}T$.   That is, the root cause of the observed Curie-Weiss growth of $\Delta f_{1/2}$ is probably related to the growth of antiferromagnetic spin correlations. 

For shorter values of $\tau$, $\Delta f_{1/2}$ begins to grow strongly below $T_{charge}$ and deviate from the Curie-Weiss behavior.  To better characterize the change of the overall lineshape due to the emergence of the wing-like signals, we also plot the half width at the 10\% signal intensity of the peak, $\Delta f_{1/10}$ (see Fig.\ 5(b) for the definition), in the inset of Fig.\ 9(a); the enhancement of $\Delta f_{1/10}$ in the charge ordered state is more pronounced than that of $\Delta f_{1/2}$.  

To understand the underlying physics of rather strong enhancement of $\Delta f_{1/2}$ and $\Delta f_{1/10}$ toward $T_{spin}^{neutron}$, it is useful to recall that the ``homogeneous linewidth'' in magnetic materials is generally enhanced by spin correlations \cite{Heller, Moriya1956, Jaccarino1965}, and hence its contribution to $\Delta f_{1/2}$ ($\propto 1/T_{2}$) generally diverges at a magnetic phase transition.  In fact, our aligned powder NMR measurements in the paramagnetic state of undoped La$_2$CuO$_{4}$ \cite{Imai1993_1, Imai1993_2} showed that the exponential growth of the spin-spin correlation length $\xi$ due to the two dimensional short range order in the renormalized classical scaling regime of the square-lattice Heisenberg model \cite{Chakravarty} leads to a strong growth of $\Delta f_{1/2}$ from 0.05~MHz at 500~K to 0.26~MHz at 440~K.  The net growth of $\Delta f_{1/2}$  from $T_{charge}$ to $T_{spin}^{neutron}$ observed for $\tau = 2~\mu$s is $\sim 0.5$~MHz, and indeed comparable with the case of La$_2$CuO$_{4}$.

\subsubsection{$^{63}{Cu}$ transverse relaxation rate $1/T_{2G}$}

In general, in the $B_{ext}~||~c$-axis geometry, one can fit the spin echo decay $M(2 \tau)$ of high $T_c$ cuprates as a convolution of the Lorentzian and Gaussian functions \cite{PenningtonPRB1989, PenningtonPRL1991,ImaiPRB1993, ItohJPSJ1992, TakigawaPRB1994};
\begin{equation}
          M(2 \tau) = M(0) \cdot exp (-\frac{2\tau}{T_{2R}}) \cdot exp (-\frac{1}{2}(\frac{2\tau}{T_{2G}})^{2}).
\label{2}
\end{equation}
$1/T_{2R}$ is the Redfield's $T_1$ contribution, and arises from the reduction of the horizontal component of the nuclear magnetization when longitudinal relaxation redirects the nuclear magnetization toward the c-axis.  $1/T_{2R}$ may be accurately estimated from the anisotropic $T_{1}$ tensor based on the Redfield theory; for the NMR central transition, 
\begin{equation}
          \frac{1}{T_{2R}} = 3 \cdot \frac{1}{T_{1}}^{(c)}+\frac{1}{T_{1}}^{(ab)},
\label{2}
\end{equation}
where the superscript c and ab represent the quantization axis set by the direction of the applied magnetic field $B_{ext}$ \cite{PenningtonPRB1989, PenningtonPRL1991}.  $\frac{1}{T_{1}}^{(c)}$ is nothing but the result presented in Fig.\ 8.  We confirmed that the anisotropy of the $T_1$ tensor is $\frac{1}{T_{1}}^{(ab)}/\frac{1}{T_{1}}^{(c)} = 3.4 \pm 0.2$, in agreement with the anisotropy $3.6 \pm 0.2$ observed in the paramagnetic state of the undoped La$_2$CuO$_{4}$ at 500~K \cite{Imai1993_1, Imai1993_2}.  

$1/T_{2G}$ represents the Gaussian component of the transverse spin-spin relaxation rate.  In high $T_c$ cuprates, $1/T_{2G}$ is much larger than the Gaussian term expected for the nuclear dipole-dipole interaction, and is dominated by the indirect nuclear spin-spin coupling through Cu electron spins in the form of $a_{ij}I_{i}^{c} \cdot I_{j}^{c}$, where $a_{ij}$ is the indirect nuclear spin-spin coupling energy \cite{PenningtonPRB1989, PenningtonPRL1991}.  In essence, the large Gaussian contribution arises from the fact that a nuclear spin $I_{i}$  located at a site $i$ precesses about a {\it static} hyperfine magnetic field $\sim a_{ij}I_{j}^{c}/\hbar \gamma_n$ induced by another nuclear spin $I_{j}$ via Cu electron spins, based on the Ruderman-Kittel mechanism \cite{PenningtonPRB1989, PenningtonPRL1991, ImaiPRB1993}.  $1/T_{2G}$ therefore reflects the real part of the wave-vector ${\bf q}$-dependent static spin susceptibility $\chi '({\bf q})$ enhanced near the antiferromagnetic wave vector ${\bf q} = (\pi/a, \pi/a)$, where $a$ is the Cu-Cu distance \cite{PenningtonPRB1989, PenningtonPRL1991, ImaiPRB1993, TakigawaPRB1994, ItohJPSJ1992}, and we expect qualitatively similar temperature dependence as $\Delta f_{1/2}$ \cite{Relation}.  

We confirmed that the Lorentzian-Gaussian convolution fit with Eq.\ (6) under the constraint on $T_{2R}$ from Eq.\ (7) is indeed good and stable near 295~K, as shown by a solid curve through the data points in Fig.\ 4.  The resulting value of  $1/T_{2G}$ changes very little even if we reduce the strength of the radio frequency pulses by a factor of two and double the pulse widths. Normally, as the Cu electron spin-spin correlation grows with decreasing temperature, $1/T_{2G}$ grows and the Gaussian curvature becomes stronger \cite{ImaiPRB1993, Imai1993_1, Imai1993_2, TakigawaPRB1994, ItohJPSJ1992}.  In the present case, however, the spin echo decay at lower temperatures becomes almost exponential without a Gaussian curvature for longer $2\tau$ above $20~\mu$s.  Analogous change in the spin echo decay near charge order transition was previously reported for La$_{1.88}$Ba$_{0.12}$CuO$_4$ \cite{TouLBCOT2} and then other striped cuprates \cite{HuntPRL, HuntPRB}.  Owing to the very short $t_{dead}$, we have been able to extend the measurement range of $M(2\tau)$ down to  $2\tau = 4~\mu$s, and found that strongly saturating behavior of $M(2\tau)$ due to the Gaussian curvature remains for very short $2\tau \lesssim 12~\mu$s even below $T_{charge}$, as shown in Fig.\ 4 and the inset of Fig.\ 10.  This trend continues down to $\sim 35$~K, below which the Gaussian curvature becomes non-existent as the spin order sets in. 

\begin{figure}
\includegraphics[width=3.5in]{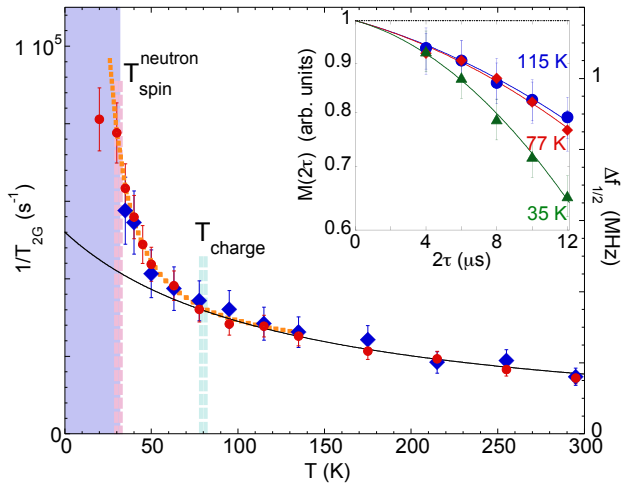}
\caption{(Main panel) Blue diamond: the Gaussian component of the spin echo decay rate, $1/T_{2G}$, obtained from the fit of $M(2\tau)$ in the short time domain below $2\tau = 12~ \mu$s.  For comparison, we also plot $\Delta f_{1/2}$ measured in the frequency domain with a fixed $\tau = 2~\mu$s (red bullet).  The solid line is the same Curie-Weiss fit above $T_{charge}$ in Fig.\ 9(a).  The dotted line through the data points below $T_{charge}$ represents a fit with the anisotropic non-linear sigma model \cite{CastronetoPRL76}, $1/T_{2G} \propto \xi \propto exp(2\pi\widetilde{\rho}_{s}/k_{B}T)$, with the effective spin stiffness $2\pi\widetilde{\rho}_{s}/k_{B} \simeq 40$~K.  (Inset) Representative semi-logarithm plots of the spin echo decay $M(2\tau)$ (normalized to $M(0)=1$ for clarity).  Solid lines: the best fit to Eq.(6) under the constraint on $T_{2R}$ set by Eq.\ (7). 
}\label{cs}
\end{figure}

In view of the fact that $\Delta f_{1/2}$ depends on $\tau$ in the charge ordered state, it may not make sense to fit the spin echo decay in the form of Eq. (6) for the entire time domain from $2\tau = 4~\mu$s to 120~$\mu$s.   Instead, we restricted the fitting range of the spin echo decay to a short time domain below  $2\tau \sim 12~\mu$s; we maintained the constraint on $1/T_{2R}$ based on Eq.(7) using the $1/T_1$ data measured with $\tau = 2~\mu$s.  We summarize $1/T_{2G}$ thus deduced in the main panel of Fig.\ 10 in comparison to $\Delta f_{1/2}$.  $1/T_{2G}$ shows nearly identical temperature dependence as $\Delta f_{1/2}$ measured at $\tau =2~\mu$s, confirming our expectation that both quantities reflect $\chi '({\bf q})$ that grows with enhanced spin-spin correlations.  We emphasize that we measured $\Delta f_{1/2}$ in the frequency domain, whereas we deduced $1/T_{2G}$ in the time domain.  

An interesting aspect of Fig.\ 10 is that $1/T_{2G}$ as well as $\Delta f_{1/2}$ exhibits a divergent trend below $T_{charge}$.  Theoretically, $1/T_{2G} \propto \xi$  \cite{Sokol}, and hence our finding signals strong growth of spin-spin correlations in the charge ordered state.  In fact, the observed temperature dependence reminds us of our earlier observation for paramagnetic La$_2$CuO$_{4}$, in which we found exponentially divergent behavior of  $1/T_{2G}$ \cite{Imai1993_2} induced by the exponential growth of $\xi \propto exp(2\pi \rho_{s}/k_{B}T)$ \cite{Chakravarty}; $2\pi \rho_{s}$ is the spin stiffness of the CuO$_2$ plane related to the Cu-Cu super-exchange interaction $J$ as $2\pi \rho_{s} = 1.13 J$  ($J/k_{B} \simeq 1500$~K for La$_2$CuO$_{4}$).  

In the present case of  La$_{1.885}$Sr$_{0.115}$CuO$_4$, the observed magnitude of $1/T_{2G} = 5.8 \times 10^{4}$~(s$^{-1}$) at 35~K is indeed comparable to $1/T_{2G} \sim 7 \times 10^{4}$ (s$^{-1}$) observed for paramagnetic La$_2$CuO$_{4}$ at 450~K with $\xi/a \sim 20$ \cite{Imai1993_1, Imai1993_2}.  We can also qualitatively account for the observed temperature dependence below $T_{charge}$ using an analogous framework based on the anisotropic non-linear sigma model with the effective spin stiffness $2\pi \widetilde{\rho}_{s}$ \cite{CastronetoPRL76} , $1/T_{2G} \propto exp(2\pi \widetilde{\rho}_{s}/k_{B}T$) \cite{Thurber3leg}; the best fit shown by a dotted curve in Fig.\ 10 resulted in $2\pi \widetilde{\rho}_{s}/k_{B} \simeq 40$~K.   

\subsubsection{$^{63}{Cu}$ NMR signal intensity wipeout}

Last but not least, we wish to address the $^{63}$Cu NMR signal intensity wipeout observed below $T_{charge}$ in Figs.\ 1 and 7, based on which the existence of charge order in the La$_2$CuO$_4$-based superconductors was originally concluded two decades ago \cite{HuntPRL, SingerPRB, HuntPRB}.  Generally, the integrated intensity of the NMR lineshape is proportional to the number of nuclear spins detected from the sample.  Accordingly, if one properly takes into account the trivial reduction of the apparent signal intensity caused by the transverse spin echo decay (such as the results in Fig.\ 4), the overall intensity should be conserved --- unless the resonant frequency shifts away or the relaxation times become too short to detect the spin echo signal.  

As shown in Fig.\ 7(a), the integrated intensity for fixed $\tau = 2~\mu$s is indeed conserved down to $\sim 40$~K, because the loss of the spectral weight of the narrower main peak below $T_{charge}$ is compensated by the growth of the wing-like signals.  As we approach $T_{spin}^{neutron}$, we begin to lose the total intensity even for $\tau = 2~\mu$s; this is because a growing fraction of the sample has extremely fast transverse relaxation in the wing-like segments due to the critical slowing down of spin fluctuations.  Below $T_{c}=30$~K, superconducting shielding effect also contributes to the the signal intensity loss, and comparison of the intensity across $T_{spin}^{neutron}$ ($=T_c$) becomes dicey.

In the case of longer $\tau = 12$ and 20~$\mu$s, the integrated intensity begins to drop precipitously below $T_{charge}$.  This is because the resonant frequency of the nuclear spins under the strong influence of charge order shifts to the wing-like segments, and their fast $T_2$ prevents them from contributing to the lineshapes for $\tau = 12$ and 20~$\mu$s.  The mild suppression of the integrated intensity that precedes from $\sim 130$~K and $\sim 200$~K for $\tau = 12$ and 20~$\mu$s, respectively, is a trivial consequence of faster spin echo decay at lower temperatures.  

We can eliminate this spin echo decay effect on the intensity by extrapolating the integrated intensity of the Gaussian lineshape observed at $\tau = 12~\mu$s (blue open squares in Fig.\ 7(a)) to $\tau =0$ using the results of spin echo decay curves $M(2\tau)$ in Fig.\ 4.  We present the extrapolated intensity, $I_{Normal}$, as black filled circles in both Fig.\ 7(a) and Fig.\ 7(b).  $I_{Normal}$ represents the net spectral weight of the normally behaving, narrower main peak without the transverse $T_2$ relaxation effect, and without the contribution of the anomalous wing-like segments under the strong influence of charge order.  By subtracting  $I_{Normal}$ from the normalized intensity, we can estimate the spectral weight of the wing-like segments as $I_{Wing} = 1-I_{Normal}$, as shown in Fig.\ 7(b).  In a separate work, we also used $^{139}$La NMR to arrive at nearly identical results as Fig.\ 7(b) \cite{Arsenault}.  

Recalling that $1/T_1$ measured for longer values of $\tau$ at the narrower main peak shows the behavior similar to the optimally superconducting La$_{1.85}$Sr$_{0.15}$CuO$_4$, $I_{Normal}$ reflects some segments of the CuO$_2$ planes that seem almost oblivious to charge order.  The volume fraction of such segments gradually diminishes below $T_{charge}$, while the volume fraction affected strongly by charge order, as represented by the spectral weight $I_{Wing}$, increases.

\section{Summary and conclusions}
We have reported a systematic $^{63}$Cu NMR investigation of the $I_{z}=+1/2$ to -1/2 central transition of a single crystal sample of La$_{1.885}$Sr$_{0.115}$CuO$_4$.  We determined $T_{charge} \simeq 80$~K of our crystal based on high precision x-ray scattering experiments \cite{He}, and compared the NMR properties above and below charge order transition.  Since the central transition depends on charge degrees of freedom only through the second order term of the EFG, $\Delta\nu_{Q}^{(2)}$, we also conducted preliminary measurements of the $I_{z}=\pm3/2$ to $\pm1/2$ satellite transitions; the latter depends on the first order effects of the EFG.  But we did not find any significant changes in the linewidth at least down to $\tau = 4~\mu$s \cite{Imai_inhomogeneity}.  This suggests that the amplitude of charge density modulation is very small in La$_{1.885}$Sr$_{0.115}$CuO$_4$, which explains why x-ray scattering experiments needed extra decade to capture the elusive Bragg scattering signals.  Our focus of the present study is therefore on the influence of charge order via enhanced spin correlations below $T_{charge}$. 

By probing the NMR properties in an extremely short time domain down to $\tau = 2~\mu$s, we demonstrated that two different types of $^{63}$Cu NMR signals exist below $T_{charge}$: a narrower main peak and extremely broad wing-like signals.  The properties of the main peak measured with longer delay times $\tau = 12 \sim 30~\mu$s are very similar to those observed for optimally superconducting La$_{1.85}$Sr$_{0.15}$CuO$_4$ even below $T_{charge}$, with canonical Curie-Weiss growth of the dynamical spin susceptibility.  On the other hand, the very broad line profile $\Delta f_{1/10}$ and the very fast relaxation times $T_1$ and $T_2$ of the wing-like segments indicate that these nuclear spins are under the strong influence of charge order that enhances spin-spin correlations.  The spectral weight $I_{Normal}$ of the normally behaving main peak is gradually wiped out below $T_{charge}$ as summarized in Fig.\ 7(b), because the lost spectral weight is progressively transferred to $I_{Wing}$ for the wing-like segments.  

The existence of  two markedly different types of domains implies that charge order does not proceed uniformly in space below $T_{charge}$.  We note that the two component nature of the CuO$_2$ planes below $T_{charge}$ manifests itself for a different measurement geometry of $B_{ext}~||~ab$-plane as well, as briefly summarized in Appendix B.   
In a separate study, we will also show that $^{139}$La NMR yields nearly identical two component picture as Fig.\ 7(b) \cite{Arsenault}.  Furthermore, the fraction of $^{139}$La NMR signals corresponding to $I_{Wing}$ reaches $\sim 100$~\% at $\sim 20$~K.  This seems to suggest that the entire volume of the CuO$_2$ planes are affected by charge order at lower temperatures, and a simple phase separation scenario seems implausible.  It is worth recalling that the charge density wave in NbSe$_2$ is known  to nucleate near the defects at much higher temperatures than the bulk transition \cite{STM}.  Analogous scenario may apply in the present case in the vicinity of, e.g., the LTO domain boundaries.  

The two component nature of the NMR lineshape observed below $T_{charge}$ indicates that a peculiar form of electronic inhomogeneity begins to develop at the onset of the charge order transition.  It remains to be seen if our finding below $T_{charge}$ is directly related to the nematic phase \cite{Kivelson} proposed for the charge ordered CuO$_2$ planes.  We note that the $^{63}$Cu NMR anomalies reported here resemble with the $^{75}$As NMR anomalies exhibited by iron-pnictides LaFeAsO \cite{Fu1111PRL} and Ba(Fe$_{1-x}$Co$_{x}$)$_2$As$_2$ \cite{TakedaPRL} when these materials undergo glassy spin order; spin nematicity is suspected in these pnictides, too.  
 
This new form of inhomogeneity that manifests itself only below $T_{charge}$ must not be confused with the pre-existing, mild inhomogeneity caused by the patch by patch variation of the local hole concentration $x_{local}$ with several nano meter length scales, associated with the quenched disorder induced by random substitution of Sr$^{2+}$.  We refer readers to Singer et al. \cite{SingerLSCOPRL, SingerPRB} for the detailed characterization of the latter.  We repeated measurements of $1/T_1$ as functions of $T$ and $\nu_{Q}$ across the upper satellite NMR transition between the nuclear spin $\pm1/2$ to $\pm3/2$ states (analogous to Fig.\ 3 in \cite{SingerLSCOPRL}), and did not find any hint of redistributions of the local hole concentration $x_{local}$ below $T_{charge}$ \cite{Imai_inhomogeneity}.  This finding is consistent with our suggestion above that a simple phase separation picture below $T_{charge}$ seems inadequate.

Our new and more comprehensive NMR results naturally explain why we were able to identify the onset of charge order of La$_2$CuO$_4$-based superconductors in our earlier $^{63}$Cu NQR work based primarily on the signal intensity wipeout effect in Fig.\ 1 \cite{HuntPRL, SingerPRB, HuntPRB}, aided by the enhancement of $1/T_1$ at $^{139}$La sites and the disappearane of the Gaussian curvature in $T_2$ spin echo decay at $^{63}$Cu sites.  In the 1990's, we were unable to access the short time domain below $\tau \sim 10~\mu$s due to the instrumental limitations set by $t_{dead} \gtrsim 10~\mu$s, and hence we did not observe the nuclear spins that belong to the wing-like segments below $T_{charge}$.  We extrapolated the integrated intensity of the NQR lineshape observed at $\tau \sim 12~\mu$s to $\tau = 0$ based on the spin echo decay curves measured for $\tau > 12~\mu$s.  Such a procedure is equivalent as our present method used to estimate $I_{Normal}$, and we have proved here that $I_{Normal}$ indeed gets wiped out precisely below $T_{charge}$ for this composition.    

We can also infer why we underestimated $T_{charge}$ of La$_{1.885}$Sr$_{0.115}$CuO$_4$ as $\sim 50$~K in our initial reports in 1999 (see Fig.\ 1(d)) \cite{HuntPRL}.  Since we were unable to observe the Gaussian curvature of $M(2\tau)$ near $T_{charge}$ that persists only for the inaccessible short time domain (Fig.\ 4 and the inset to Fig.\ 10), we overestimated the integrated intensity between $\sim$50~K and $T_{charge}$ by extrapolating $M(2\tau)$ to $\tau =0$ by incorrectly assuming a purely exponential form below $\tau \sim 12~\mu$s.  

The superconducting phase transition of La$_{1.885}$Sr$_{0.115}$CuO$_4$ sets in at $T_{c}=30$~K simultaneously as the onset of spin order at $T_{spin}^{neutron} = 30$~K \cite{Kimura}.  In that context, it is important to notice that the spectral weight $I_{Normal}$ of the normally behaving, narrower main peak still accounts for nearly $\sim 1/3$ of the volume fraction of the CuO$_2$ planes at 30~K.  The flip side of this observation is that the volume fraction of the truly static spin order as observed by $\mu$SR measurements reaches only $\sim 20$~\% \cite{Savici}.  The continuing debate over the coexistence of superconductivity and magnetism in the CuO$_2$ planes must take these observations into consideration.  

From the temperature dependence of $1/T_{2G}$ analyzed with the anisotropic non-linear sigma model, we also estimated the effective spin stiffness of the charge ordered state as $2\pi \widetilde{\rho}_{s}/k_{B} \simeq 40$~K, in comparison to $2\pi \widetilde{\rho}_{s}/k_{B} \simeq 200$~K in La$_{1.48}$Nd$_{0.4}$Sr$_{0.12}$CuO$_4$ \cite{Tranquada} and La$_{1.68}$Eu$_{0.2}$Sr$_{0.12}$CuO$_4$ \cite{HuntPRB}.  The value for La$_{1.885}$Sr$_{0.115}$CuO$_4$ may be somewhat underestimated, because we deduced it from the linewidth of the main peak disregarding the wing-like segments (if we fit $\Delta f_{1/10}$ to the same form, we obtain $2\pi \widetilde{\rho}_{s}/k_{B} \simeq 50$~K); in addition, our fit is not conducted in the low temperature limit. We note that Mitrovic et al. arrived at an even smaller value $2\pi \widetilde{\rho}_{s}/k_{B} \simeq 25$~K for La$_{1.88}$Sr$_{0.12}$CuO$_4$ by fitting the temperature dependence of $1/T_1$ below 40~K, but they did not consider charge order as the driving mechanism behind the magnetic anomalies they observed below 80~K \cite{Mitrovic}.  

Regardless, the small value of the effective spin stiffness is consistent with the fact that the onset of the spin ordering $T_{spin}^{neutron} \simeq 30$~K in La$_{1.885}$Sr$_{0.115}$CuO$_4$  \cite{Kimura} is much lower than $T_{spin}^{neutron} \simeq 50$~K observed for La$_{1.48}$Nd$_{0.4}$Sr$_{0.12}$CuO$_4$ \cite{Tranquada}.   It is also consistent with our earlier finding that the Zeeman perturbed $^{63}$Cu NQR signal was barely observable even at 0.35~K due to the residual dynamics of spins \cite{SingerLT23}, whereas the static nature of the hyperfine magnetic field at 0.35~K in La$_{1.88}$Ba$_{0.12}$CuO$_4$ and La$_{1.68}$Eu$_{0.2}$Sr$_{0.12}$CuO$_4$  resulted in nearly full recovery of the integrated intensity, as shown in Fig.\ 1(b-c). 

The charge ordered state realized in the LTO structure of La$_{1.885}$Sr$_{0.115}$CuO$_4$ may have other unique aspects, too.  $1/T_1$ of a majority of nuclear spins belonging to the narrower main peak still decreases below $T_{charge}$; this implies that a large volume fraction of CuO$_2$ plane is still far from magnetic instability.  In contrast, in both La$_{1.88}$Ba$_{0.12}$CuO$_4$ and La$_{1.68}$Eu$_{0.2}$Sr$_{0.12}$CuO$_4$, $1/T_1$ measured at the $^{63}$Cu sites \cite{TouLBCOT1, HuntPRB, Pelc} and $^{139}$La sites \cite{HuntPRB, BaekLaT1PRB2015} begin to diverge below $T_{charge}$. Perhaps magnetic correlations develop more uniformly in space in the charge ordered state realized in the LTT structure.  Furthermore, the linewidth of the nearest-neighbor $^{63}$Cu sites of Ba$^{2+}$ in La$_{1.88}$Ba$_{0.12}$CuO$_4$ measured with NQR using $\tau = 2~\mu$s near $\sim 40$~MHz has recently been reported to broaden as much as 70\% in the charge ordered state \cite{Pelc}, but so far we have not found such a dramatic effect at least down to $\tau = 4~\mu$s \cite{Imai_inhomogeneity}.  This may be another indication that the amplitude of the charge density modulation in La$_{1.885}$Sr$_{0.115}$CuO$_4$ is much smaller than in La$_{1.88}$Ba$_{0.12}$CuO$_4$.     One should be cautioned, however, that our earlier work in the paramagnetic state of ${\it undoped}$ La$_{2}$CuO$_4$ found that both $^{63}$Cu NQR and NMR linewidth show a divergent behavior below 700~K as the spin-spin correlations grow exponentially \cite{Imai1993_1, Imai1993_2}.  In other words, $^{63}$Cu NQR linewidth can grow simply due to the enhanced spin-spin correlations via indirect nuclear spin couplings.  In La$_{1.88}$Ba$_{0.12}$CuO$_4$, divergent behavior of  $1/T_1$ below $T_{charge} \sim 54$~K indicates that spin correlations are indeed quickly growing toward $T_{spin}^{neutron}$.  Therefore, it is not clear if the observed NQR line broadening \cite{Pelc} reflects the distribution of $\nu_{Q}$ caused by the charge density modulation, or simply the magnetic correlation effects analogous to the case of La$_{2}$CuO$_4$.    

Another open question is the origine of charge localization and its relation to NMR anomalies for $x < 0.1$.   In the case of $x \sim 1/8$, earlier charge transport measurements by Komiya and Ando on La$_{1.88}$Sr$_{0.12}$CuO$_4$ in high magnetic fields up to 60~T \cite{Ando} demonstrated that charge localization is driven by nothing but charge order.  In fact, the characteristic temperature they dubbed as the localization temperature, $T_{loc} \simeq 80$~K, agrees with $T_{charge}$.  Moreover, $T_{loc}$ shows a local maximum at $x \sim 1/8$ in their phase diagram,  in agreement with $T_{charge}$ \cite{Croft}.  But the situation is more complex below $x \sim 0.1$, because $T_{loc}$ \cite{Ando} increases for lower doping, whereas $T_{charge}$ decreases \cite{Croft}.  Related to this issue, our initial identification of the on-set of the Cu signal intensity wipeout as the onset of charge order  below $x \sim 0.1$ in  La$_{2-x}$Sr$_{x}$CuO$_4$ with or without Nd co-doping \cite{HuntPRL,SingerPRB} later turned out to be false \cite{HuntPRB}; comparison with $T_{charge}$ determined subsequently by neutron scattering revealed that the onset of wipeout in the low doping regime is more closely related to the charge localization; instead, it was the inflection point in the temperature dependence of the signal intensity wipeout that should have been identified as $T_{charge}$ (see Fig. 18 in \cite{HuntPRB} where we summarized the characteristic temperature scale of resistivity upturn, inflection point in the wipeout, and $T_{charge}$ as determined by neutron).  From very early days, it had been known that paramagnetic $^{63}$Cu NMR signal intensity gradually disappears \cite{Imai1990, Imai1993_1} and NMR linewidth and relaxation rates begin to grow \cite{Thurber2122} when the in-plane resistivity deviates from the linear temperature dependence in the lightly doped region.  Intuitively, this can be easily understood: when the mobility of some holes is lost, the spin-spin correlations  would locally grow in their neighborhood, leading to extremely fast NMR relaxation rates and a broad line profile.  In the case of $x=0.06$, earlier high field NMR work found a broad line similar to what we reported here \cite{Julien1999PRL}.  In view of the fact that charge order Bragg peaks were finally observed near $x \sim 1/8$ \cite{He, Croft, Thampy}, it will be interesting to investigate charge order and the nature of localization with NMR in the broader range of composition below $x \sim1/8$ with fresh eyes.

\section{Acknowledgement}
T. I. would like to express his sincere gratitude to his early collaborators on the NMR investigation of charge order in La$_{1.885}$Sr$_{0.115}$CuO$_4$, in particular Dr.\ Allen W.\ Hunt, Dr.\ Philip M.\ Singer, Dr.\ Kent R.\ Thurber, and Dr. Agneta F. Cederst\"{o}m.  They conducted foundational NMR research on charge order in the La$_2$CuO$_4$-based superconductors during the late 1990's, and their pioneering discoveries formed the basis of the present work.  The work at McMaster was financially supported by NSERC and CIFAR.  The work at Stanford was supported by US Department of Energy, Office of Science, Office of Basic Energy Sciences under grant no. DE-FG02- 07ER46134.  The work at Tohoku was supported by Grant-in-Aid for Scientific Research (A) (16H02125).

\section{Appendix A: Examples of $1/T_{1}$ recovery curves}
In Fig.\ 11, we show representative recovery curves $M_{T_{1}}(t)$ after an inversion $\pi$ pulse observed for $1/T_{1}$ measurements at 40~K under various conditions.  The solid curves represent the best fit to the theoretical expression for the magnetic transition between the $I_{z} = 1/2$ to $-1/2$ states \cite{Narath}, 
\begin{equation}
          M_{T_{1}}(t) = A - B (\frac{9}{10}e^{-6t/T_{1}}+\frac{1}{10}e^{-t/T_{1}}),
\label{8}
\end{equation}
where $A$ and $B$ represent the saturated and inverted intensity, respectively.  $A$, $B$ and $1/T_{1}$ are the free fitting parameters.  For the ease of comparison, we normalized the intensity in Fig.\ 11.   The fit is satisfactory.   Notice that the recovery curve measured for the wing-like signal at 104.13~MHz with $\tau = 2~\mu$s is much faster than at the main peak.  
\begin{figure}
\includegraphics[width=3.5in]{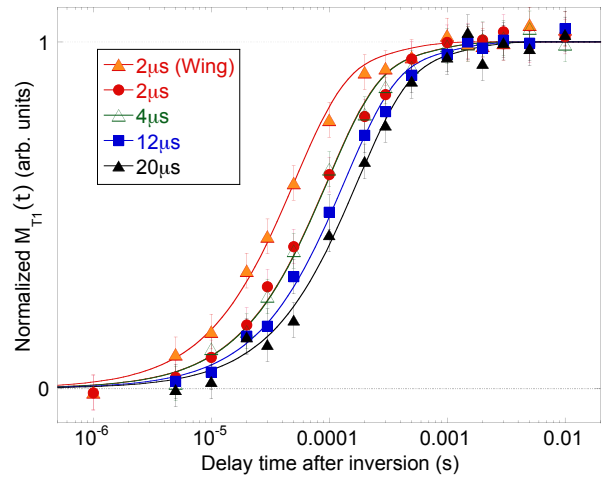}
\caption{Representative $^{63}$Cu NMR $1/T_{1}$ recovery curves of the $I_{z} =+1/2$ to $-1/2$ central transition observed at 40~K for different values of $\tau = 2~\mu$s, $4~\mu$s, $12~\mu$s, and $20~\mu$s at the normal peak.  The solid curves are the best fit with Eq.\ 8.  Also shown is the recovery curve for the wing-like signal observed at 104.13~MHz.   
}\label{cs}
\end{figure}

\section{Appendix B: Measurements with the $B_{ext}~||~ab$ geometry}

\begin{figure}
\includegraphics[width=3in]{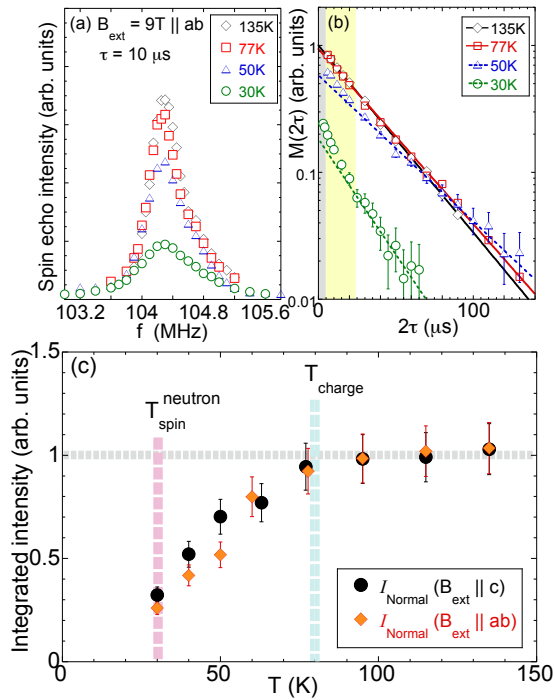}
\caption{(a) Representative $^{63}$Cu NMR lineshapes of the $I_{z} =+1/2$ to $-1/2$ central transition observed with $\tau = 10~\mu$s in $B_{ext}=9T~||~ab$.  The signal intensity is normalized for the Boltzman factor by multiplying temperature $T$.  (b) The corresponding spin echo decay observed at the peak.  Solid lines through the 135~K and 77~K data are the best exponential fit, whereas the dashed lines through the 50~K and 30~K data represent the best exponential fit above $2\tau = 20~\mu$s.  (c) The temperature dependence of $I_{Normal}$ deduced from (a) and (b) for  $B_{ext}~||~ab$ (diamond) agrees very well with the result for $B_{ext}~||~c$ (filled bullets, from Fig.\ 7). 
}\label{cs}
\end{figure}

In Fig.\ 12(a), we show representative $^{63}$Cu NMR lineshapes measured in $B_{ext}=9$~T applied along the ab-plane.  In this field geometry, the peak frequency is shifted to $\sim 104.3$~MHz primarily by the second order quadrupole term in Eq.(1), $\Delta \nu_{Q}^{(2)} \sim \frac{3 \nu_{Q}^{2}}{16 \gamma_{n}B_{ext}} \sim 2.3$~MHz, where we used the nuclear quadrupole frequency $\nu_{Q} \sim 35$~MHz for the majority $^{63}$Cu A-sites \cite{SingerLSCOPRL}.  Since the $^{63}$Cu B-sites nearest-neighbor to Sr$^{2+}$ ions have a larger $\nu_{Q} \sim 38$~MHz \cite{SingerLSCOPRL}, the lineshape is somewhat asymmetrical, with a hump on the higher frequency side.  The broad NMR linewidth is set primarily by the two different values of $\nu_{Q}$ and their large distributions even below $T_{charge}$.

We present the spin echo decay curves in Fig.\ 12(b).  Above $T_{charge}$, the spin echo decay is purely Lorentzian, because the Gaussian term in Eq.\ (6), caused by the indirect nuclear spin-spin coupling effect, is motionally narrowed to an exponential in this field geometry \cite{PenningtonPRB1989}.   It also means that the indirect nuclear spin-spin coupling is ineffective in magnetic line broadening.  The two component nature of $^{63}$Cu NMR signals between those arising from the segments affected strongly by charge order (corresponding to the wing-like segments) and those from the rest of the CuO$_2$ planes manifests itself in the spin echo decay curves below $T_{charge}$; notice that the spin echo decay  measured at 50~K and 30~K is no longer single exponential, and exhibits a quick initial decay up to $2\tau \sim 20~\mu$s.   

We can estimate the spectral weight $I_{Normal}$ of the normally behaving segments of CuO$_2$ planes below $T_{charge}$ by extrapolating the spin echo decay curves observed above $2\tau = 20~\mu$s exponentially to $2\tau =0$, as shown by the dashed lines in Fig.\ 12(b).  We summarize $I_{Normal}$ thus deduced for $B_{ext}~||~ab$ in Fig.\ 12(c), in comparison to the result for $B_{ext}~||~c$.  The agreement  is very good.  \\

\pagebreak


\begin{thebibliography}{71}%
\makeatletter
\providecommand \@ifxundefined [1]{%
 \@ifx{#1\undefined}
}%
\providecommand \@ifnum [1]{%
 \ifnum #1\expandafter \@firstoftwo
 \else \expandafter \@secondoftwo
 \fi
}%
\providecommand \@ifx [1]{%
 \ifx #1\expandafter \@firstoftwo
 \else \expandafter \@secondoftwo
 \fi
}%
\providecommand \natexlab [1]{#1}%
\providecommand \enquote  [1]{``#1''}%
\providecommand \bibnamefont  [1]{#1}%
\providecommand \bibfnamefont [1]{#1}%
\providecommand \citenamefont [1]{#1}%
\providecommand \href@noop [0]{\@secondoftwo}%
\providecommand \href [0]{\begingroup \@sanitize@url \@href}%
\providecommand \@href[1]{\@@startlink{#1}\@@href}%
\providecommand \@@href[1]{\endgroup#1\@@endlink}%
\providecommand \@sanitize@url [0]{\catcode `\\12\catcode `\$12\catcode
  `\&12\catcode `\#12\catcode `\^12\catcode `\_12\catcode `\%12\relax}%
\providecommand \@@startlink[1]{}%
\providecommand \@@endlink[0]{}%
\providecommand \url  [0]{\begingroup\@sanitize@url \@url }%
\providecommand \@url [1]{\endgroup\@href {#1}{\urlprefix }}%
\providecommand \urlprefix  [0]{URL }%
\providecommand \Eprint [0]{\href }%
\providecommand \doibase [0]{http://dx.doi.org/}%
\providecommand \selectlanguage [0]{\@gobble}%
\providecommand \bibinfo  [0]{\@secondoftwo}%
\providecommand \bibfield  [0]{\@secondoftwo}%
\providecommand \translation [1]{[#1]}%
\providecommand \BibitemOpen [0]{}%
\providecommand \bibitemStop [0]{}%
\providecommand \bibitemNoStop [0]{.\EOS\space}%
\providecommand \EOS [0]{\spacefactor3000\relax}%
\providecommand \BibitemShut  [1]{\csname bibitem#1\endcsname}%
\let\auto@bib@innerbib\@empty
\bibitem [{\citenamefont {Axe}\ \emph {et~al.}(1989)\citenamefont {Axe},
  \citenamefont {Moudden}, \citenamefont {Hohlwein}, \citenamefont {Cox},
  \citenamefont {Mohanty}, \citenamefont {Moodenbaugh},\ and\ \citenamefont
  {Xu}}]{Axe}%
  \BibitemOpen
  \bibfield  {author} {\bibinfo {author} {\bibfnamefont {J.~D.}\ \bibnamefont
  {Axe}}, \bibinfo {author} {\bibfnamefont {A.~H.}\ \bibnamefont {Moudden}},
  \bibinfo {author} {\bibfnamefont {D.}~\bibnamefont {Hohlwein}}, \bibinfo
  {author} {\bibfnamefont {D.~E.}\ \bibnamefont {Cox}}, \bibinfo {author}
  {\bibfnamefont {K.~M.}\ \bibnamefont {Mohanty}}, \bibinfo {author}
  {\bibfnamefont {A.~R.}\ \bibnamefont {Moodenbaugh}}, \ and\ \bibinfo {author}
  {\bibfnamefont {Y.}~\bibnamefont {Xu}},\ }\href {\doibase
  10.1103/PhysRevLett.62.2751} {\bibfield  {journal} {\bibinfo  {journal}
  {Phys. Rev. Lett.}\ }\textbf {\bibinfo {volume} {62}},\ \bibinfo {pages}
  {2751} (\bibinfo {year} {1989})}\BibitemShut {NoStop}%
\bibitem [{\citenamefont {Crawford}\ \emph {et~al.}(1991)\citenamefont
  {Crawford}, \citenamefont {Harlow}, \citenamefont {McCarron}, \citenamefont
  {Farneth}, \citenamefont {Axe}, \citenamefont {Chou},\ and\ \citenamefont
  {Huang}}]{Crawford}%
  \BibitemOpen
  \bibfield  {author} {\bibinfo {author} {\bibfnamefont {M.~K.}\ \bibnamefont
  {Crawford}}, \bibinfo {author} {\bibfnamefont {R.~L.}\ \bibnamefont
  {Harlow}}, \bibinfo {author} {\bibfnamefont {E.~M.}\ \bibnamefont
  {McCarron}}, \bibinfo {author} {\bibfnamefont {W.~E.}\ \bibnamefont
  {Farneth}}, \bibinfo {author} {\bibfnamefont {J.~D.}\ \bibnamefont {Axe}},
  \bibinfo {author} {\bibfnamefont {H.}~\bibnamefont {Chou}}, \ and\ \bibinfo
  {author} {\bibfnamefont {Q.}~\bibnamefont {Huang}},\ }\href {\doibase
  10.1103/PhysRevB.44.7749} {\bibfield  {journal} {\bibinfo  {journal} {Phys.
  Rev. B}\ }\textbf {\bibinfo {volume} {44}},\ \bibinfo {pages} {7749}
  (\bibinfo {year} {1991})}\BibitemShut {NoStop}%
\bibitem [{\citenamefont {Luke}\ \emph {et~al.}(1991)\citenamefont {Luke},
  \citenamefont {Le}, \citenamefont {Strenlieb}, \citenamefont {Wu},
  \citenamefont {Uemura}, \citenamefont {Brewer},\ and\ \citenamefont
  {Riseman}}]{Luke}%
  \BibitemOpen
  \bibfield  {author} {\bibinfo {author} {\bibfnamefont {G.~M.}\ \bibnamefont
  {Luke}}, \bibinfo {author} {\bibfnamefont {L.~P.}\ \bibnamefont {Le}},
  \bibinfo {author} {\bibfnamefont {B.~J.}\ \bibnamefont {Strenlieb}}, \bibinfo
  {author} {\bibfnamefont {W.~D.}\ \bibnamefont {Wu}}, \bibinfo {author}
  {\bibfnamefont {Y.~J.}\ \bibnamefont {Uemura}}, \bibinfo {author}
  {\bibfnamefont {J.~H.}\ \bibnamefont {Brewer}}, \ and\ \bibinfo {author}
  {\bibnamefont {Riseman}},\ }\href@noop {} {\bibfield  {journal} {\bibinfo
  {journal} {Physica C}\ }\textbf {\bibinfo {volume} {185-189}},\ \bibinfo
  {pages} {1175} (\bibinfo {year} {1991})}\BibitemShut {NoStop}%
\bibitem [{\citenamefont {Nachumi}\ \emph {et~al.}(1998)\citenamefont
  {Nachumi}, \citenamefont {Fudamoto}, \citenamefont {Keren}, \citenamefont
  {Kojima}, \citenamefont {Larkin}, \citenamefont {Luke}, \citenamefont
  {Merrin}, \citenamefont {Tchernyshyov}, \citenamefont {Uemura}, \citenamefont
  {Ichikawa}, \citenamefont {Goto}, \citenamefont {Takagi}, \citenamefont
  {Uchida}, \citenamefont {Crawford}, \citenamefont {McCarron}, \citenamefont
  {MacLaughlin},\ and\ \citenamefont {Heffner}}]{Nachumi}%
  \BibitemOpen
  \bibfield  {author} {\bibinfo {author} {\bibfnamefont {B.}~\bibnamefont
  {Nachumi}}, \bibinfo {author} {\bibfnamefont {Y.}~\bibnamefont {Fudamoto}},
  \bibinfo {author} {\bibfnamefont {A.}~\bibnamefont {Keren}}, \bibinfo
  {author} {\bibfnamefont {K.~M.}\ \bibnamefont {Kojima}}, \bibinfo {author}
  {\bibfnamefont {M.}~\bibnamefont {Larkin}}, \bibinfo {author} {\bibfnamefont
  {G.~M.}\ \bibnamefont {Luke}}, \bibinfo {author} {\bibfnamefont
  {J.}~\bibnamefont {Merrin}}, \bibinfo {author} {\bibfnamefont
  {O.}~\bibnamefont {Tchernyshyov}}, \bibinfo {author} {\bibfnamefont {Y.~J.}\
  \bibnamefont {Uemura}}, \bibinfo {author} {\bibfnamefont {N.}~\bibnamefont
  {Ichikawa}}, \bibinfo {author} {\bibfnamefont {M.}~\bibnamefont {Goto}},
  \bibinfo {author} {\bibfnamefont {H.}~\bibnamefont {Takagi}}, \bibinfo
  {author} {\bibfnamefont {S.}~\bibnamefont {Uchida}}, \bibinfo {author}
  {\bibfnamefont {M.~K.}\ \bibnamefont {Crawford}}, \bibinfo {author}
  {\bibfnamefont {E.~M.}\ \bibnamefont {McCarron}}, \bibinfo {author}
  {\bibfnamefont {D.~E.}\ \bibnamefont {MacLaughlin}}, \ and\ \bibinfo {author}
  {\bibfnamefont {R.~H.}\ \bibnamefont {Heffner}},\ }\href {\doibase
  10.1103/PhysRevB.58.8760} {\bibfield  {journal} {\bibinfo  {journal} {Phys.
  Rev. B}\ }\textbf {\bibinfo {volume} {58}},\ \bibinfo {pages} {8760}
  (\bibinfo {year} {1998})}\BibitemShut {NoStop}%
\bibitem [{\citenamefont {Tranquada}\ \emph {et~al.}(1995)\citenamefont
  {Tranquada}, \citenamefont {Sternlieb}, \citenamefont {Axe}, \citenamefont
  {Nakamura},\ and\ \citenamefont {Uchida}}]{Tranquada}%
  \BibitemOpen
  \bibfield  {author} {\bibinfo {author} {\bibfnamefont {J.~M.}\ \bibnamefont
  {Tranquada}}, \bibinfo {author} {\bibfnamefont {B.~J.}\ \bibnamefont
  {Sternlieb}}, \bibinfo {author} {\bibfnamefont {J.~D.}\ \bibnamefont {Axe}},
  \bibinfo {author} {\bibfnamefont {Y.}~\bibnamefont {Nakamura}}, \ and\
  \bibinfo {author} {\bibfnamefont {S.}~\bibnamefont {Uchida}},\ }\href@noop {}
  {\bibfield  {journal} {\bibinfo  {journal} {Nature}\ }\textbf {\bibinfo
  {volume} {375}},\ \bibinfo {pages} {561} (\bibinfo {year}
  {1995})}\BibitemShut {NoStop}%
\bibitem [{\citenamefont {Tranquada}\ \emph {et~al.}(1996)\citenamefont
  {Tranquada}, \citenamefont {Axe}, \citenamefont {Ichikawa}, \citenamefont
  {Nakamura}, \citenamefont {Uchida},\ and\ \citenamefont
  {Nachumi}}]{TranquadaPRB54}%
  \BibitemOpen
  \bibfield  {author} {\bibinfo {author} {\bibfnamefont {J.~M.}\ \bibnamefont
  {Tranquada}}, \bibinfo {author} {\bibfnamefont {J.~D.}\ \bibnamefont {Axe}},
  \bibinfo {author} {\bibfnamefont {N.}~\bibnamefont {Ichikawa}}, \bibinfo
  {author} {\bibfnamefont {Y.}~\bibnamefont {Nakamura}}, \bibinfo {author}
  {\bibfnamefont {S.}~\bibnamefont {Uchida}}, \ and\ \bibinfo {author}
  {\bibfnamefont {B.}~\bibnamefont {Nachumi}},\ }\href {\doibase
  10.1103/PhysRevB.54.7489} {\bibfield  {journal} {\bibinfo  {journal} {Phys.
  Rev. B}\ }\textbf {\bibinfo {volume} {54}},\ \bibinfo {pages} {7489}
  (\bibinfo {year} {1996})}\BibitemShut {NoStop}%
\bibitem [{\citenamefont {Tranquada}\ \emph {et~al.}(1999)\citenamefont
  {Tranquada}, \citenamefont {Ichikawa},\ and\ \citenamefont
  {Uchida}}]{TranquadaPRB59}%
  \BibitemOpen
  \bibfield  {author} {\bibinfo {author} {\bibfnamefont {J.~M.}\ \bibnamefont
  {Tranquada}}, \bibinfo {author} {\bibfnamefont {N.}~\bibnamefont {Ichikawa}},
  \ and\ \bibinfo {author} {\bibfnamefont {S.}~\bibnamefont {Uchida}},\ }\href
  {\doibase 10.1103/PhysRevB.59.14712} {\bibfield  {journal} {\bibinfo
  {journal} {Phys. Rev. B}\ }\textbf {\bibinfo {volume} {59}},\ \bibinfo
  {pages} {14712} (\bibinfo {year} {1999})}\BibitemShut {NoStop}%
\bibitem [{\citenamefont {Imai}\ \emph {et~al.}(1990)\citenamefont {Imai},
  \citenamefont {Yoshimura}, \citenamefont {Uemura}, \citenamefont {Yasuoka},\
  and\ \citenamefont {Kosuge}}]{Imai1990}%
  \BibitemOpen
  \bibfield  {author} {\bibinfo {author} {\bibfnamefont {T.}~\bibnamefont
  {Imai}}, \bibinfo {author} {\bibfnamefont {K.}~\bibnamefont {Yoshimura}},
  \bibinfo {author} {\bibfnamefont {T.}~\bibnamefont {Uemura}}, \bibinfo
  {author} {\bibfnamefont {H.}~\bibnamefont {Yasuoka}}, \ and\ \bibinfo
  {author} {\bibfnamefont {K.}~\bibnamefont {Kosuge}},\ }\href@noop {}
  {\bibfield  {journal} {\bibinfo  {journal} {J. Phys. Soc. Jpn.}\ }\textbf
  {\bibinfo {volume} {59}},\ \bibinfo {pages} {3846} (\bibinfo {year}
  {1990})}\BibitemShut {NoStop}%
\bibitem [{\citenamefont {Hunt}\ \emph {et~al.}(1999)\citenamefont {Hunt},
  \citenamefont {Singer}, \citenamefont {Thurber},\ and\ \citenamefont
  {Imai}}]{HuntPRL}%
  \BibitemOpen
  \bibfield  {author} {\bibinfo {author} {\bibfnamefont {A.~W.}\ \bibnamefont
  {Hunt}}, \bibinfo {author} {\bibfnamefont {P.~M.}\ \bibnamefont {Singer}},
  \bibinfo {author} {\bibfnamefont {K.~R.}\ \bibnamefont {Thurber}}, \ and\
  \bibinfo {author} {\bibfnamefont {T.}~\bibnamefont {Imai}},\ }\href {\doibase
  10.1103/PhysRevLett.82.4300} {\bibfield  {journal} {\bibinfo  {journal}
  {Phys. Rev. Lett.}\ }\textbf {\bibinfo {volume} {82}},\ \bibinfo {pages}
  {4300} (\bibinfo {year} {1999})}\BibitemShut {NoStop}%
\bibitem [{\citenamefont {Singer}\ \emph {et~al.}(1999)\citenamefont {Singer},
  \citenamefont {Hunt}, \citenamefont {Cederstr\"om},\ and\ \citenamefont
  {Imai}}]{SingerPRB}%
  \BibitemOpen
  \bibfield  {author} {\bibinfo {author} {\bibfnamefont {P.~M.}\ \bibnamefont
  {Singer}}, \bibinfo {author} {\bibfnamefont {A.~W.}\ \bibnamefont {Hunt}},
  \bibinfo {author} {\bibfnamefont {A.~F.}\ \bibnamefont {Cederstr\"om}}, \
  and\ \bibinfo {author} {\bibfnamefont {T.}~\bibnamefont {Imai}},\ }\href
  {\doibase 10.1103/PhysRevB.60.15345} {\bibfield  {journal} {\bibinfo
  {journal} {Phys. Rev. B}\ }\textbf {\bibinfo {volume} {60}},\ \bibinfo
  {pages} {15345} (\bibinfo {year} {1999})}\BibitemShut {NoStop}%
\bibitem [{\citenamefont {Hunt}\ \emph {et~al.}(2001)\citenamefont {Hunt},
  \citenamefont {Singer}, \citenamefont {Cederstr\"om},\ and\ \citenamefont
  {Imai}}]{HuntPRB}%
  \BibitemOpen
  \bibfield  {author} {\bibinfo {author} {\bibfnamefont {A.~W.}\ \bibnamefont
  {Hunt}}, \bibinfo {author} {\bibfnamefont {P.~M.}\ \bibnamefont {Singer}},
  \bibinfo {author} {\bibfnamefont {A.~F.}\ \bibnamefont {Cederstr\"om}}, \
  and\ \bibinfo {author} {\bibfnamefont {T.}~\bibnamefont {Imai}},\ }\href
  {\doibase 10.1103/PhysRevB.64.134525} {\bibfield  {journal} {\bibinfo
  {journal} {Phys. Rev. B}\ }\textbf {\bibinfo {volume} {64}},\ \bibinfo
  {pages} {134525} (\bibinfo {year} {2001})}\BibitemShut {NoStop}%
\bibitem [{\citenamefont {Cederstr\"om}\ \emph {et~al.}(2001)\citenamefont
  {Cederstr\"om}, \citenamefont {Hunt}, \citenamefont {Singer},\ and\
  \citenamefont {Imai}}]{Cederstrom}%
  \BibitemOpen
  \bibfield  {author} {\bibinfo {author} {\bibfnamefont {A.~F.}\ \bibnamefont
  {Cederstr\"om}}, \bibinfo {author} {\bibfnamefont {A.~W.}\ \bibnamefont
  {Hunt}}, \bibinfo {author} {\bibfnamefont {P.~M.}\ \bibnamefont {Singer}}, \
  and\ \bibinfo {author} {\bibfnamefont {T.}~\bibnamefont {Imai}},\ }\href@noop
  {} {\bibfield  {journal} {\bibinfo  {journal} {J. Magn. Magn. Mater.}\
  }\textbf {\bibinfo {volume} {226-230}},\ \bibinfo {pages} {863} (\bibinfo
  {year} {2001})}\BibitemShut {NoStop}%
\bibitem [{\citenamefont {Singer}\ \emph {et~al.}(2003)\citenamefont {Singer},
  \citenamefont {Hunt}, \citenamefont {Cederstr\"om},\ and\ \citenamefont
  {Imai}}]{SingerLT23}%
  \BibitemOpen
  \bibfield  {author} {\bibinfo {author} {\bibfnamefont {P.~M.}\ \bibnamefont
  {Singer}}, \bibinfo {author} {\bibfnamefont {A.~W.}\ \bibnamefont {Hunt}},
  \bibinfo {author} {\bibfnamefont {A.~F.}\ \bibnamefont {Cederstr\"om}}, \
  and\ \bibinfo {author} {\bibfnamefont {T.}~\bibnamefont {Imai}},\ }\href@noop
  {} {\bibfield  {journal} {\bibinfo  {journal} {Physica C}\ }\textbf {\bibinfo
  {volume} {388-389}},\ \bibinfo {pages} {209} (\bibinfo {year}
  {2003})}\BibitemShut {NoStop}%
\bibitem [{\citenamefont {Imai}\ and\ \citenamefont {Hirota}()}]{ImaiAspen}%
  \BibitemOpen
  \bibfield  {author} {\bibinfo {author} {\bibfnamefont {T.}~\bibnamefont
  {Imai}}\ and\ \bibinfo {author} {\bibfnamefont {K.}~\bibnamefont {Hirota}},\
  }\href@noop {} {}\bibinfo {note} {Unpublished $^{63}$Cu, $^{139}$La, and
  $^{17}$O NMR work on a La$_{1.885}$Sr$_{0.115}$CuO$_{4}$ single crystal
  presented first at the Aspen Winter Conference on Quantum Criticality
  (January 1999)}\BibitemShut {NoStop}%
\bibitem [{\citenamefont {Service}(1999)}]{ServiceScience}%
  \BibitemOpen
  \bibfield  {author} {\bibinfo {author} {\bibfnamefont {R.}~\bibnamefont
  {Service}},\ }\href@noop {} {\bibfield  {journal} {\bibinfo  {journal}
  {Science}\ }\textbf {\bibinfo {volume} {283}},\ \bibinfo {pages} {1116 }
  (\bibinfo {year} {1999})}\BibitemShut {NoStop}%
\bibitem [{\citenamefont {Abu-Shiekah}\ \emph {et~al.}(1999)\citenamefont
  {Abu-Shiekah}, \citenamefont {Bernal}, \citenamefont {Menovsky},
  \citenamefont {Brom},\ and\ \citenamefont {Zaanen}}]{Abu-Shiekah1999}%
  \BibitemOpen
  \bibfield  {author} {\bibinfo {author} {\bibfnamefont {I.~M.}\ \bibnamefont
  {Abu-Shiekah}}, \bibinfo {author} {\bibfnamefont {O.~O.}\ \bibnamefont
  {Bernal}}, \bibinfo {author} {\bibfnamefont {A.~A.}\ \bibnamefont
  {Menovsky}}, \bibinfo {author} {\bibfnamefont {H.~B.}\ \bibnamefont {Brom}},
  \ and\ \bibinfo {author} {\bibfnamefont {J.}~\bibnamefont {Zaanen}},\ }\href
  {\doibase 10.1103/PhysRevLett.83.3309} {\bibfield  {journal} {\bibinfo
  {journal} {Phys. Rev. Lett.}\ }\textbf {\bibinfo {volume} {83}},\ \bibinfo
  {pages} {3309} (\bibinfo {year} {1999})}\BibitemShut {NoStop}%
\bibitem [{\citenamefont {Fujita}\ \emph {et~al.}(2004)\citenamefont {Fujita},
  \citenamefont {Goka}, \citenamefont {Yamada}, \citenamefont {Tranquada},\
  and\ \citenamefont {Regnault}}]{Fujita}%
  \BibitemOpen
  \bibfield  {author} {\bibinfo {author} {\bibfnamefont {M.}~\bibnamefont
  {Fujita}}, \bibinfo {author} {\bibfnamefont {H.}~\bibnamefont {Goka}},
  \bibinfo {author} {\bibfnamefont {K.}~\bibnamefont {Yamada}}, \bibinfo
  {author} {\bibfnamefont {J.~M.}\ \bibnamefont {Tranquada}}, \ and\ \bibinfo
  {author} {\bibfnamefont {L.~P.}\ \bibnamefont {Regnault}},\ }\href {\doibase
  10.1103/PhysRevB.70.104517} {\bibfield  {journal} {\bibinfo  {journal} {Phys.
  Rev. B}\ }\textbf {\bibinfo {volume} {70}},\ \bibinfo {pages} {104517}
  (\bibinfo {year} {2004})}\BibitemShut {NoStop}%
\bibitem [{\citenamefont {Fink}\ \emph {et~al.}(2011)\citenamefont {Fink},
  \citenamefont {Soltwisch}, \citenamefont {Geck}, \citenamefont {Schierle},
  \citenamefont {Weschke},\ and\ \citenamefont {B\"uchner}}]{Fink}%
  \BibitemOpen
  \bibfield  {author} {\bibinfo {author} {\bibfnamefont {J.}~\bibnamefont
  {Fink}}, \bibinfo {author} {\bibfnamefont {V.}~\bibnamefont {Soltwisch}},
  \bibinfo {author} {\bibfnamefont {J.}~\bibnamefont {Geck}}, \bibinfo {author}
  {\bibfnamefont {E.}~\bibnamefont {Schierle}}, \bibinfo {author}
  {\bibfnamefont {E.}~\bibnamefont {Weschke}}, \ and\ \bibinfo {author}
  {\bibfnamefont {B.}~\bibnamefont {B\"uchner}},\ }\href {\doibase
  10.1103/PhysRevB.83.092503} {\bibfield  {journal} {\bibinfo  {journal} {Phys.
  Rev. B}\ }\textbf {\bibinfo {volume} {83}},\ \bibinfo {pages} {092503}
  (\bibinfo {year} {2011})}\BibitemShut {NoStop}%
\bibitem [{\citenamefont {Fink}\ \emph {et~al.}(2009)\citenamefont {Fink},
  \citenamefont {Schierle}, \citenamefont {Weschke}, \citenamefont {Geck},
  \citenamefont {Hawthorn}, \citenamefont {Soltwisch}, \citenamefont {Wadati},
  \citenamefont {Wu}, \citenamefont {D\"urr}, \citenamefont {Wizent},
  \citenamefont {B\"uchner},\ and\ \citenamefont {Sawatzky}}]{Fink2}%
  \BibitemOpen
  \bibfield  {author} {\bibinfo {author} {\bibfnamefont {J.}~\bibnamefont
  {Fink}}, \bibinfo {author} {\bibfnamefont {E.}~\bibnamefont {Schierle}},
  \bibinfo {author} {\bibfnamefont {E.}~\bibnamefont {Weschke}}, \bibinfo
  {author} {\bibfnamefont {J.}~\bibnamefont {Geck}}, \bibinfo {author}
  {\bibfnamefont {D.}~\bibnamefont {Hawthorn}}, \bibinfo {author}
  {\bibfnamefont {V.}~\bibnamefont {Soltwisch}}, \bibinfo {author}
  {\bibfnamefont {H.}~\bibnamefont {Wadati}}, \bibinfo {author} {\bibfnamefont
  {H.-H.}\ \bibnamefont {Wu}}, \bibinfo {author} {\bibfnamefont {H.~A.}\
  \bibnamefont {D\"urr}}, \bibinfo {author} {\bibfnamefont {N.}~\bibnamefont
  {Wizent}}, \bibinfo {author} {\bibfnamefont {B.}~\bibnamefont {B\"uchner}}, \
  and\ \bibinfo {author} {\bibfnamefont {G.~A.}\ \bibnamefont {Sawatzky}},\
  }\href {\doibase 10.1103/PhysRevB.79.100502} {\bibfield  {journal} {\bibinfo
  {journal} {Phys. Rev. B}\ }\textbf {\bibinfo {volume} {79}},\ \bibinfo
  {pages} {100502} (\bibinfo {year} {2009})}\BibitemShut {NoStop}%
\bibitem [{\citenamefont {He}\ \emph {et~al.}(2017)\citenamefont {He},
  \citenamefont {Lee},\ and\ \citenamefont {Fujita}}]{He}%
  \BibitemOpen
  \bibfield  {author} {\bibinfo {author} {\bibfnamefont {W.}~\bibnamefont
  {He}}, \bibinfo {author} {\bibfnamefont {Y.~S.}\ \bibnamefont {Lee}}, \ and\
  \bibinfo {author} {\bibfnamefont {M.}~\bibnamefont {Fujita}},\ }\href@noop {}
  {} (\bibinfo {year} {2017}),\ \bibinfo {note} {unpublished}\BibitemShut
  {NoStop}%
\bibitem [{\citenamefont {Thampy}\ \emph {et~al.}(2014)\citenamefont {Thampy},
  \citenamefont {Dean}, \citenamefont {Christensen}, \citenamefont {Steinke},
  \citenamefont {Islam}, \citenamefont {Oda}, \citenamefont {Ido},
  \citenamefont {Momono}, \citenamefont {Wilkins},\ and\ \citenamefont
  {Hill}}]{Thampy}%
  \BibitemOpen
  \bibfield  {author} {\bibinfo {author} {\bibfnamefont {V.}~\bibnamefont
  {Thampy}}, \bibinfo {author} {\bibfnamefont {M.~P.~M.}\ \bibnamefont {Dean}},
  \bibinfo {author} {\bibfnamefont {N.~B.}\ \bibnamefont {Christensen}},
  \bibinfo {author} {\bibfnamefont {L.}~\bibnamefont {Steinke}}, \bibinfo
  {author} {\bibfnamefont {Z.}~\bibnamefont {Islam}}, \bibinfo {author}
  {\bibfnamefont {M.}~\bibnamefont {Oda}}, \bibinfo {author} {\bibfnamefont
  {M.}~\bibnamefont {Ido}}, \bibinfo {author} {\bibfnamefont {N.}~\bibnamefont
  {Momono}}, \bibinfo {author} {\bibfnamefont {S.~B.}\ \bibnamefont {Wilkins}},
  \ and\ \bibinfo {author} {\bibfnamefont {J.~P.}\ \bibnamefont {Hill}},\
  }\href {\doibase 10.1103/PhysRevB.90.100510} {\bibfield  {journal} {\bibinfo
  {journal} {Phys. Rev. B}\ }\textbf {\bibinfo {volume} {90}},\ \bibinfo
  {pages} {100510} (\bibinfo {year} {2014})}\BibitemShut {NoStop}%
\bibitem [{\citenamefont {Croft}\ \emph {et~al.}(2014)\citenamefont {Croft},
  \citenamefont {Lester}, \citenamefont {Senn}, \citenamefont {Bombardi},\ and\
  \citenamefont {Hayden}}]{Croft}%
  \BibitemOpen
  \bibfield  {author} {\bibinfo {author} {\bibfnamefont {T.~P.}\ \bibnamefont
  {Croft}}, \bibinfo {author} {\bibfnamefont {C.}~\bibnamefont {Lester}},
  \bibinfo {author} {\bibfnamefont {M.~S.}\ \bibnamefont {Senn}}, \bibinfo
  {author} {\bibfnamefont {A.}~\bibnamefont {Bombardi}}, \ and\ \bibinfo
  {author} {\bibfnamefont {S.~M.}\ \bibnamefont {Hayden}},\ }\href {\doibase
  10.1103/PhysRevB.89.224513} {\bibfield  {journal} {\bibinfo  {journal} {Phys.
  Rev. B}\ }\textbf {\bibinfo {volume} {89}},\ \bibinfo {pages} {224513}
  (\bibinfo {year} {2014})}\BibitemShut {NoStop}%
\bibitem [{\citenamefont {Kimura}\ \emph {et~al.}(1999)\citenamefont {Kimura},
  \citenamefont {Hirota}, \citenamefont {Matsushita}, \citenamefont {Yamada},
  \citenamefont {Endoh}, \citenamefont {Lee}, \citenamefont {Majkrzak},
  \citenamefont {Erwin}, \citenamefont {Shirane}, \citenamefont {Greven},
  \citenamefont {Lee}, \citenamefont {Kastner},\ and\ \citenamefont
  {Birgeneau}}]{Kimura}%
  \BibitemOpen
  \bibfield  {author} {\bibinfo {author} {\bibfnamefont {H.}~\bibnamefont
  {Kimura}}, \bibinfo {author} {\bibfnamefont {K.}~\bibnamefont {Hirota}},
  \bibinfo {author} {\bibfnamefont {H.}~\bibnamefont {Matsushita}}, \bibinfo
  {author} {\bibfnamefont {K.}~\bibnamefont {Yamada}}, \bibinfo {author}
  {\bibfnamefont {Y.}~\bibnamefont {Endoh}}, \bibinfo {author} {\bibfnamefont
  {S.-H.}\ \bibnamefont {Lee}}, \bibinfo {author} {\bibfnamefont {C.~F.}\
  \bibnamefont {Majkrzak}}, \bibinfo {author} {\bibfnamefont {R.}~\bibnamefont
  {Erwin}}, \bibinfo {author} {\bibfnamefont {G.}~\bibnamefont {Shirane}},
  \bibinfo {author} {\bibfnamefont {M.}~\bibnamefont {Greven}}, \bibinfo
  {author} {\bibfnamefont {Y.~S.}\ \bibnamefont {Lee}}, \bibinfo {author}
  {\bibfnamefont {M.~A.}\ \bibnamefont {Kastner}}, \ and\ \bibinfo {author}
  {\bibfnamefont {R.~J.}\ \bibnamefont {Birgeneau}},\ }\href {\doibase
  10.1103/PhysRevB.59.6517} {\bibfield  {journal} {\bibinfo  {journal} {Phys.
  Rev. B}\ }\textbf {\bibinfo {volume} {59}},\ \bibinfo {pages} {6517}
  (\bibinfo {year} {1999})}\BibitemShut {NoStop}%
\bibitem [{\citenamefont {Suzuki}\ \emph {et~al.}(1999)\citenamefont {Suzuki},
  \citenamefont {Oshima}, \citenamefont {Chiba}, \citenamefont {Fukase},
  \citenamefont {Goto}, \citenamefont {Kimura},\ and\ \citenamefont
  {Yamada}}]{SuzukiResistivity}%
  \BibitemOpen
  \bibfield  {author} {\bibinfo {author} {\bibfnamefont {T.}~\bibnamefont
  {Suzuki}}, \bibinfo {author} {\bibfnamefont {Y.}~\bibnamefont {Oshima}},
  \bibinfo {author} {\bibfnamefont {K.}~\bibnamefont {Chiba}}, \bibinfo
  {author} {\bibfnamefont {T.}~\bibnamefont {Fukase}}, \bibinfo {author}
  {\bibfnamefont {T.}~\bibnamefont {Goto}}, \bibinfo {author} {\bibfnamefont
  {H.}~\bibnamefont {Kimura}}, \ and\ \bibinfo {author} {\bibfnamefont
  {K.}~\bibnamefont {Yamada}},\ }\href {\doibase 10.1103/PhysRevB.60.10500}
  {\bibfield  {journal} {\bibinfo  {journal} {Phys. Rev. B}\ }\textbf {\bibinfo
  {volume} {60}},\ \bibinfo {pages} {10500} (\bibinfo {year}
  {1999})}\BibitemShut {NoStop}%
\bibitem [{\citenamefont {Takeda}\ \emph {et~al.}(2014)\citenamefont {Takeda},
  \citenamefont {Imai}, \citenamefont {Tachibana}, \citenamefont {Gaudet},
  \citenamefont {Gaulin}, \citenamefont {Saparov},\ and\ \citenamefont
  {Sefat}}]{TakedaPRL}%
  \BibitemOpen
  \bibfield  {author} {\bibinfo {author} {\bibfnamefont {H.}~\bibnamefont
  {Takeda}}, \bibinfo {author} {\bibfnamefont {T.}~\bibnamefont {Imai}},
  \bibinfo {author} {\bibfnamefont {M.}~\bibnamefont {Tachibana}}, \bibinfo
  {author} {\bibfnamefont {J.}~\bibnamefont {Gaudet}}, \bibinfo {author}
  {\bibfnamefont {B.~D.}\ \bibnamefont {Gaulin}}, \bibinfo {author}
  {\bibfnamefont {B.~I.}\ \bibnamefont {Saparov}}, \ and\ \bibinfo {author}
  {\bibfnamefont {A.~S.}\ \bibnamefont {Sefat}},\ }\href {\doibase
  10.1103/PhysRevLett.113.117001} {\bibfield  {journal} {\bibinfo  {journal}
  {Phys. Rev. Lett.}\ }\textbf {\bibinfo {volume} {113}},\ \bibinfo {pages}
  {117001} (\bibinfo {year} {2014})}\BibitemShut {NoStop}%
\bibitem [{\citenamefont {Slichter}(1990)}]{Slichter}%
  \BibitemOpen
  \bibfield  {author} {\bibinfo {author} {\bibfnamefont {C.~P.}\ \bibnamefont
  {Slichter}},\ }\href@noop {} {\emph {\bibinfo {title} {Principles of magnetic
  resonance}}},\ \bibinfo {edition} {3rd}\ ed.\ (\bibinfo  {publisher}
  {Springer-Verlag},\ \bibinfo {address} {Berlin},\ \bibinfo {year}
  {1990})\BibitemShut {NoStop}%
\bibitem [{\citenamefont {Mila}\ and\ \citenamefont {Rice}(1989)}]{Mila}%
  \BibitemOpen
  \bibfield  {author} {\bibinfo {author} {\bibfnamefont {F.}~\bibnamefont
  {Mila}}\ and\ \bibinfo {author} {\bibfnamefont {T.~M.}\ \bibnamefont
  {Rice}},\ }\href@noop {} {\bibfield  {journal} {\bibinfo  {journal} {Physica
  C}\ }\textbf {\bibinfo {volume} {157}},\ \bibinfo {pages} {561} (\bibinfo
  {year} {1989})}\BibitemShut {NoStop}%
\bibitem [{\citenamefont {Takigawa}\ \emph
  {et~al.}(1989{\natexlab{a}})\citenamefont {Takigawa}, \citenamefont {Hammel},
  \citenamefont {Heffner},\ and\ \citenamefont {Fisk}}]{TakigawaPRB1989shift}%
  \BibitemOpen
  \bibfield  {author} {\bibinfo {author} {\bibfnamefont {M.}~\bibnamefont
  {Takigawa}}, \bibinfo {author} {\bibfnamefont {P.~C.}\ \bibnamefont
  {Hammel}}, \bibinfo {author} {\bibfnamefont {R.~H.}\ \bibnamefont {Heffner}},
  \ and\ \bibinfo {author} {\bibfnamefont {Z.}~\bibnamefont {Fisk}},\ }\href
  {\doibase 10.1103/PhysRevB.39.7371} {\bibfield  {journal} {\bibinfo
  {journal} {Phys. Rev. B}\ }\textbf {\bibinfo {volume} {39}},\ \bibinfo
  {pages} {7371} (\bibinfo {year} {1989}{\natexlab{a}})}\BibitemShut {NoStop}%
\bibitem [{\citenamefont {Barrett}\ \emph {et~al.}(1990)\citenamefont
  {Barrett}, \citenamefont {Durand}, \citenamefont {Pennington}, \citenamefont
  {Slichter}, \citenamefont {Friedmann}, \citenamefont {Rice},\ and\
  \citenamefont {Ginsberg}}]{Barrett}%
  \BibitemOpen
  \bibfield  {author} {\bibinfo {author} {\bibfnamefont {S.~E.}\ \bibnamefont
  {Barrett}}, \bibinfo {author} {\bibfnamefont {D.~J.}\ \bibnamefont {Durand}},
  \bibinfo {author} {\bibfnamefont {C.~H.}\ \bibnamefont {Pennington}},
  \bibinfo {author} {\bibfnamefont {C.~P.}\ \bibnamefont {Slichter}}, \bibinfo
  {author} {\bibfnamefont {T.~A.}\ \bibnamefont {Friedmann}}, \bibinfo {author}
  {\bibfnamefont {J.~P.}\ \bibnamefont {Rice}}, \ and\ \bibinfo {author}
  {\bibfnamefont {D.~M.}\ \bibnamefont {Ginsberg}},\ }\href {\doibase
  10.1103/PhysRevB.41.6283} {\bibfield  {journal} {\bibinfo  {journal} {Phys.
  Rev. B}\ }\textbf {\bibinfo {volume} {41}},\ \bibinfo {pages} {6283}
  (\bibinfo {year} {1990})}\BibitemShut {NoStop}%
\bibitem [{\citenamefont {Imai}\ \emph
  {et~al.}(1993{\natexlab{a}})\citenamefont {Imai}, \citenamefont {Slichter},
  \citenamefont {Yoshimura},\ and\ \citenamefont {Kosuge}}]{Imai1993_1}%
  \BibitemOpen
  \bibfield  {author} {\bibinfo {author} {\bibfnamefont {T.}~\bibnamefont
  {Imai}}, \bibinfo {author} {\bibfnamefont {C.~P.}\ \bibnamefont {Slichter}},
  \bibinfo {author} {\bibfnamefont {K.}~\bibnamefont {Yoshimura}}, \ and\
  \bibinfo {author} {\bibfnamefont {K.}~\bibnamefont {Kosuge}},\ }\href
  {\doibase 10.1103/PhysRevLett.70.1002} {\bibfield  {journal} {\bibinfo
  {journal} {Phys. Rev. Lett.}\ }\textbf {\bibinfo {volume} {70}},\ \bibinfo
  {pages} {1002} (\bibinfo {year} {1993}{\natexlab{a}})}\BibitemShut {NoStop}%
\bibitem [{\citenamefont {Imai}\ \emph
  {et~al.}(1993{\natexlab{b}})\citenamefont {Imai}, \citenamefont {Slichter},
  \citenamefont {Yoshimura}, \citenamefont {Katoh},\ and\ \citenamefont
  {Kosuge}}]{Imai1993_2}%
  \BibitemOpen
  \bibfield  {author} {\bibinfo {author} {\bibfnamefont {T.}~\bibnamefont
  {Imai}}, \bibinfo {author} {\bibfnamefont {C.~P.}\ \bibnamefont {Slichter}},
  \bibinfo {author} {\bibfnamefont {K.}~\bibnamefont {Yoshimura}}, \bibinfo
  {author} {\bibfnamefont {M.}~\bibnamefont {Katoh}}, \ and\ \bibinfo {author}
  {\bibfnamefont {K.}~\bibnamefont {Kosuge}},\ }\href {\doibase
  10.1103/PhysRevLett.71.1254} {\bibfield  {journal} {\bibinfo  {journal}
  {Phys. Rev. Lett.}\ }\textbf {\bibinfo {volume} {71}},\ \bibinfo {pages}
  {1254} (\bibinfo {year} {1993}{\natexlab{b}})}\BibitemShut {NoStop}%
\bibitem [{\citenamefont {Johnston}(1989)}]{Johnston}%
  \BibitemOpen
  \bibfield  {author} {\bibinfo {author} {\bibfnamefont {D.~C.}\ \bibnamefont
  {Johnston}},\ }\href {\doibase 10.1103/PhysRevLett.62.957} {\bibfield
  {journal} {\bibinfo  {journal} {Phys. Rev. Lett.}\ }\textbf {\bibinfo
  {volume} {62}},\ \bibinfo {pages} {957} (\bibinfo {year} {1989})}\BibitemShut
  {NoStop}%
\bibitem [{\citenamefont {Ishida}\ \emph {et~al.}(1989)\citenamefont {Ishida},
  \citenamefont {Kitaoka},\ and\ \citenamefont {Asayama}}]{Ishida}%
  \BibitemOpen
  \bibfield  {author} {\bibinfo {author} {\bibfnamefont {K.}~\bibnamefont
  {Ishida}}, \bibinfo {author} {\bibfnamefont {Y.}~\bibnamefont {Kitaoka}}, \
  and\ \bibinfo {author} {\bibfnamefont {K.}~\bibnamefont {Asayama}},\
  }\href@noop {} {\bibfield  {journal} {\bibinfo  {journal} {J. Phys. Soc.
  Jpn.}\ }\textbf {\bibinfo {volume} {58}},\ \bibinfo {pages} {36} (\bibinfo
  {year} {1989})}\BibitemShut {NoStop}%
\bibitem [{\citenamefont {Ohsugi}\ \emph {et~al.}(1994)\citenamefont {Ohsugi},
  \citenamefont {Kitaoka}, \citenamefont {Ishida}, \citenamefont {G.-q.},\ and\
  \citenamefont {K.}}]{Ohsugi1994}%
  \BibitemOpen
  \bibfield  {author} {\bibinfo {author} {\bibfnamefont {S.}~\bibnamefont
  {Ohsugi}}, \bibinfo {author} {\bibfnamefont {Y.}~\bibnamefont {Kitaoka}},
  \bibinfo {author} {\bibfnamefont {K.}~\bibnamefont {Ishida}}, \bibinfo
  {author} {\bibfnamefont {Z.}~\bibnamefont {G.-q.}}, \ and\ \bibinfo {author}
  {\bibfnamefont {A.}~\bibnamefont {K.}},\ }\href@noop {} {\bibfield  {journal}
  {\bibinfo  {journal} {J. Phys. Soc. Jpn.}\ }\textbf {\bibinfo {volume}
  {63}},\ \bibinfo {pages} {700} (\bibinfo {year} {1994})}\BibitemShut
  {NoStop}%
\bibitem [{\citenamefont {Imai}\ \emph {et~al.}(1988)\citenamefont {Imai},
  \citenamefont {Shimizu}, \citenamefont {Yasuoka}, \citenamefont {Ueda},\ and\
  \citenamefont {Kosuge}}]{Imai1988}%
  \BibitemOpen
  \bibfield  {author} {\bibinfo {author} {\bibfnamefont {T.}~\bibnamefont
  {Imai}}, \bibinfo {author} {\bibfnamefont {T.}~\bibnamefont {Shimizu}},
  \bibinfo {author} {\bibfnamefont {H.}~\bibnamefont {Yasuoka}}, \bibinfo
  {author} {\bibfnamefont {Y.}~\bibnamefont {Ueda}}, \ and\ \bibinfo {author}
  {\bibfnamefont {K.}~\bibnamefont {Kosuge}},\ }\href@noop {} {\bibfield
  {journal} {\bibinfo  {journal} {J. Phys. Soc. Jpn.}\ }\textbf {\bibinfo
  {volume} {57}},\ \bibinfo {pages} {2280} (\bibinfo {year}
  {1988})}\BibitemShut {NoStop}%
\bibitem [{\citenamefont {Moriya}(1963)}]{Moriya1963}%
  \BibitemOpen
  \bibfield  {author} {\bibinfo {author} {\bibfnamefont {T.}~\bibnamefont
  {Moriya}},\ }\href@noop {} {\bibfield  {journal} {\bibinfo  {journal} {J.
  Phys. Soc. Jpn.}\ }\textbf {\bibinfo {volume} {18}},\ \bibinfo {pages} {516}
  (\bibinfo {year} {1963})}\BibitemShut {NoStop}%
\bibitem [{\citenamefont {Millis}\ \emph {et~al.}(1990)\citenamefont {Millis},
  \citenamefont {Monien},\ and\ \citenamefont {Pines}}]{MMP1990}%
  \BibitemOpen
  \bibfield  {author} {\bibinfo {author} {\bibfnamefont {A.~J.}\ \bibnamefont
  {Millis}}, \bibinfo {author} {\bibfnamefont {H.}~\bibnamefont {Monien}}, \
  and\ \bibinfo {author} {\bibfnamefont {D.}~\bibnamefont {Pines}},\ }\href
  {\doibase 10.1103/PhysRevB.42.167} {\bibfield  {journal} {\bibinfo  {journal}
  {Phys. Rev. B}\ }\textbf {\bibinfo {volume} {42}},\ \bibinfo {pages} {167}
  (\bibinfo {year} {1990})}\BibitemShut {NoStop}%
\bibitem [{\citenamefont {R\o{}mer}\ \emph {et~al.}(2013)\citenamefont
  {R\o{}mer}, \citenamefont {Chang}, \citenamefont {Christensen}, \citenamefont
  {Andersen}, \citenamefont {Lefmann}, \citenamefont {M\"ahler}, \citenamefont
  {Gavilano}, \citenamefont {Gilardi}, \citenamefont {Niedermayer},
  \citenamefont {R\o{}nnow}, \citenamefont {Schneidewind}, \citenamefont
  {Link}, \citenamefont {Oda}, \citenamefont {Ido}, \citenamefont {Momono},\
  and\ \citenamefont {Mesot}}]{RomerNeutron}%
  \BibitemOpen
  \bibfield  {author} {\bibinfo {author} {\bibfnamefont {A.~T.}\ \bibnamefont
  {R\o{}mer}}, \bibinfo {author} {\bibfnamefont {J.}~\bibnamefont {Chang}},
  \bibinfo {author} {\bibfnamefont {N.~B.}\ \bibnamefont {Christensen}},
  \bibinfo {author} {\bibfnamefont {B.~M.}\ \bibnamefont {Andersen}}, \bibinfo
  {author} {\bibfnamefont {K.}~\bibnamefont {Lefmann}}, \bibinfo {author}
  {\bibfnamefont {L.}~\bibnamefont {M\"ahler}}, \bibinfo {author}
  {\bibfnamefont {J.}~\bibnamefont {Gavilano}}, \bibinfo {author}
  {\bibfnamefont {R.}~\bibnamefont {Gilardi}}, \bibinfo {author} {\bibfnamefont
  {C.}~\bibnamefont {Niedermayer}}, \bibinfo {author} {\bibfnamefont {H.~M.}\
  \bibnamefont {R\o{}nnow}}, \bibinfo {author} {\bibfnamefont {A.}~\bibnamefont
  {Schneidewind}}, \bibinfo {author} {\bibfnamefont {P.}~\bibnamefont {Link}},
  \bibinfo {author} {\bibfnamefont {M.}~\bibnamefont {Oda}}, \bibinfo {author}
  {\bibfnamefont {M.}~\bibnamefont {Ido}}, \bibinfo {author} {\bibfnamefont
  {N.}~\bibnamefont {Momono}}, \ and\ \bibinfo {author} {\bibfnamefont
  {J.}~\bibnamefont {Mesot}},\ }\href {\doibase 10.1103/PhysRevB.87.144513}
  {\bibfield  {journal} {\bibinfo  {journal} {Phys. Rev. B}\ }\textbf {\bibinfo
  {volume} {87}},\ \bibinfo {pages} {144513} (\bibinfo {year}
  {2013})}\BibitemShut {NoStop}%
\bibitem [{\citenamefont {Arsenault~et. al.}(shed)}]{Arsenault}%
  \BibitemOpen
  \bibfield  {author} {\bibinfo {author} {\bibfnamefont {A.}~\bibnamefont
  {Arsenault~et. al.}},\ }\href@noop {} {} (\bibinfo {year} {To be
  published})\BibitemShut {NoStop}%
\bibitem [{\citenamefont {Mitrovi\ifmmode~\acute{c}\else \'{c}\fi{}}\ \emph
  {et~al.}(2008)\citenamefont {Mitrovi\ifmmode~\acute{c}\else \'{c}\fi{}},
  \citenamefont {Julien}, \citenamefont {de~Vaulx}, \citenamefont
  {Horvati\ifmmode~\acute{c}\else \'{c}\fi{}}, \citenamefont {Berthier},
  \citenamefont {Suzuki},\ and\ \citenamefont {Yamada}}]{Mitrovic}%
  \BibitemOpen
  \bibfield  {author} {\bibinfo {author} {\bibfnamefont {V.~F.}\ \bibnamefont
  {Mitrovi\ifmmode~\acute{c}\else \'{c}\fi{}}}, \bibinfo {author}
  {\bibfnamefont {M.-H.}\ \bibnamefont {Julien}}, \bibinfo {author}
  {\bibfnamefont {C.}~\bibnamefont {de~Vaulx}}, \bibinfo {author}
  {\bibfnamefont {M.}~\bibnamefont {Horvati\ifmmode~\acute{c}\else
  \'{c}\fi{}}}, \bibinfo {author} {\bibfnamefont {C.}~\bibnamefont {Berthier}},
  \bibinfo {author} {\bibfnamefont {T.}~\bibnamefont {Suzuki}}, \ and\ \bibinfo
  {author} {\bibfnamefont {K.}~\bibnamefont {Yamada}},\ }\href {\doibase
  10.1103/PhysRevB.78.014504} {\bibfield  {journal} {\bibinfo  {journal} {Phys.
  Rev. B}\ }\textbf {\bibinfo {volume} {78}},\ \bibinfo {pages} {014504}
  (\bibinfo {year} {2008})}\BibitemShut {NoStop}%
\bibitem [{\citenamefont {Singer}\ \emph {et~al.}(2002)\citenamefont {Singer},
  \citenamefont {Hunt},\ and\ \citenamefont {Imai}}]{SingerLSCOPRL}%
  \BibitemOpen
  \bibfield  {author} {\bibinfo {author} {\bibfnamefont {P.~M.}\ \bibnamefont
  {Singer}}, \bibinfo {author} {\bibfnamefont {A.~W.}\ \bibnamefont {Hunt}}, \
  and\ \bibinfo {author} {\bibfnamefont {T.}~\bibnamefont {Imai}},\ }\href
  {\doibase 10.1103/PhysRevLett.88.047602} {\bibfield  {journal} {\bibinfo
  {journal} {Phys. Rev. Lett.}\ }\textbf {\bibinfo {volume} {88}},\ \bibinfo
  {pages} {047602} (\bibinfo {year} {2002})}\BibitemShut {NoStop}%
\bibitem [{\citenamefont {Singer}\ \emph {et~al.}(2005)\citenamefont {Singer},
  \citenamefont {Imai}, \citenamefont {Chou}, \citenamefont {Hirota},
  \citenamefont {Takaba}, \citenamefont {Kakeshita}, \citenamefont {Eisaki},\
  and\ \citenamefont {Uchida}}]{SingerPRB2005}%
  \BibitemOpen
  \bibfield  {author} {\bibinfo {author} {\bibfnamefont {P.~M.}\ \bibnamefont
  {Singer}}, \bibinfo {author} {\bibfnamefont {T.}~\bibnamefont {Imai}},
  \bibinfo {author} {\bibfnamefont {F.~C.}\ \bibnamefont {Chou}}, \bibinfo
  {author} {\bibfnamefont {K.}~\bibnamefont {Hirota}}, \bibinfo {author}
  {\bibfnamefont {M.}~\bibnamefont {Takaba}}, \bibinfo {author} {\bibfnamefont
  {T.}~\bibnamefont {Kakeshita}}, \bibinfo {author} {\bibfnamefont
  {H.}~\bibnamefont {Eisaki}}, \ and\ \bibinfo {author} {\bibfnamefont
  {S.}~\bibnamefont {Uchida}},\ }\href {\doibase 10.1103/PhysRevB.72.014537}
  {\bibfield  {journal} {\bibinfo  {journal} {Phys. Rev. B}\ }\textbf {\bibinfo
  {volume} {72}},\ \bibinfo {pages} {014537} (\bibinfo {year}
  {2005})}\BibitemShut {NoStop}%
\bibitem [{\citenamefont {Kumagai}\ \emph {et~al.}(1994)\citenamefont
  {Kumagai}, \citenamefont {Kawano}, \citenamefont {Watanabe}, \citenamefont
  {Nishiyama},\ and\ \citenamefont {Nagamine}}]{Kumagai}%
  \BibitemOpen
  \bibfield  {author} {\bibinfo {author} {\bibfnamefont {K.}~\bibnamefont
  {Kumagai}}, \bibinfo {author} {\bibfnamefont {K.}~\bibnamefont {Kawano}},
  \bibinfo {author} {\bibfnamefont {I.}~\bibnamefont {Watanabe}}, \bibinfo
  {author} {\bibfnamefont {K.}~\bibnamefont {Nishiyama}}, \ and\ \bibinfo
  {author} {\bibfnamefont {K.}~\bibnamefont {Nagamine}},\ }\href@noop {}
  {\bibfield  {journal} {\bibinfo  {journal} {Hyperfine Interactions}\ }\textbf
  {\bibinfo {volume} {86}},\ \bibinfo {pages} {473} (\bibinfo {year}
  {1994})}\BibitemShut {NoStop}%
\bibitem [{\citenamefont {Savici}\ \emph {et~al.}(2002)\citenamefont {Savici},
  \citenamefont {Fudamoto}, \citenamefont {Gat}, \citenamefont {Ito},
  \citenamefont {Larkin}, \citenamefont {Uemura}, \citenamefont {Luke},
  \citenamefont {Kojima}, \citenamefont {Lee}, \citenamefont {Kastner},
  \citenamefont {Birgeneau},\ and\ \citenamefont {Yamada}}]{Savici}%
  \BibitemOpen
  \bibfield  {author} {\bibinfo {author} {\bibfnamefont {A.~T.}\ \bibnamefont
  {Savici}}, \bibinfo {author} {\bibfnamefont {Y.}~\bibnamefont {Fudamoto}},
  \bibinfo {author} {\bibfnamefont {I.~M.}\ \bibnamefont {Gat}}, \bibinfo
  {author} {\bibfnamefont {T.}~\bibnamefont {Ito}}, \bibinfo {author}
  {\bibfnamefont {M.~I.}\ \bibnamefont {Larkin}}, \bibinfo {author}
  {\bibfnamefont {Y.~J.}\ \bibnamefont {Uemura}}, \bibinfo {author}
  {\bibfnamefont {G.~M.}\ \bibnamefont {Luke}}, \bibinfo {author}
  {\bibfnamefont {K.~M.}\ \bibnamefont {Kojima}}, \bibinfo {author}
  {\bibfnamefont {Y.~S.}\ \bibnamefont {Lee}}, \bibinfo {author} {\bibfnamefont
  {M.~A.}\ \bibnamefont {Kastner}}, \bibinfo {author} {\bibfnamefont {R.~J.}\
  \bibnamefont {Birgeneau}}, \ and\ \bibinfo {author} {\bibfnamefont
  {K.}~\bibnamefont {Yamada}},\ }\href {\doibase 10.1103/PhysRevB.66.014524}
  {\bibfield  {journal} {\bibinfo  {journal} {Phys. Rev. B}\ }\textbf {\bibinfo
  {volume} {66}},\ \bibinfo {pages} {014524} (\bibinfo {year}
  {2002})}\BibitemShut {NoStop}%
\bibitem [{\citenamefont {Takigawa}\ \emph
  {et~al.}(1989{\natexlab{b}})\citenamefont {Takigawa}, \citenamefont {Hammel},
  \citenamefont {Heffner}, \citenamefont {Fisk}, \citenamefont {Smith},\ and\
  \citenamefont {Schwarz}}]{TakigawaPRB39}%
  \BibitemOpen
  \bibfield  {author} {\bibinfo {author} {\bibfnamefont {M.}~\bibnamefont
  {Takigawa}}, \bibinfo {author} {\bibfnamefont {P.~C.}\ \bibnamefont
  {Hammel}}, \bibinfo {author} {\bibfnamefont {R.~H.}\ \bibnamefont {Heffner}},
  \bibinfo {author} {\bibfnamefont {Z.}~\bibnamefont {Fisk}}, \bibinfo {author}
  {\bibfnamefont {J.~L.}\ \bibnamefont {Smith}}, \ and\ \bibinfo {author}
  {\bibfnamefont {R.~B.}\ \bibnamefont {Schwarz}},\ }\href {\doibase
  10.1103/PhysRevB.39.300} {\bibfield  {journal} {\bibinfo  {journal} {Phys.
  Rev. B}\ }\textbf {\bibinfo {volume} {39}},\ \bibinfo {pages} {300} (\bibinfo
  {year} {1989}{\natexlab{b}})}\BibitemShut {NoStop}%
\bibitem [{\citenamefont {Haase}\ \emph {et~al.}(2002)\citenamefont {Haase},
  \citenamefont {Slichter},\ and\ \citenamefont {T.}}]{Haase}%
  \BibitemOpen
  \bibfield  {author} {\bibinfo {author} {\bibfnamefont {J.}~\bibnamefont
  {Haase}}, \bibinfo {author} {\bibfnamefont {C.~P.}\ \bibnamefont {Slichter}},
  \ and\ \bibinfo {author} {\bibfnamefont {C.~J.}\ \bibnamefont {Milling}},\
  }\href@noop {} {\bibfield  {journal} {\bibinfo  {journal} {J. of
  Superconductivity}\ }\textbf {\bibinfo {volume} {15}},\ \bibinfo {pages}
  {339} (\bibinfo {year} {2002})}\BibitemShut {NoStop}%
\bibitem [{\citenamefont {Heller}\ and\ \citenamefont
  {Benedek}(1962)}]{Heller}%
  \BibitemOpen
  \bibfield  {author} {\bibinfo {author} {\bibfnamefont {P.}~\bibnamefont
  {Heller}}\ and\ \bibinfo {author} {\bibfnamefont {G.~B.}\ \bibnamefont
  {Benedek}},\ }\href {\doibase 10.1103/PhysRevLett.8.428} {\bibfield
  {journal} {\bibinfo  {journal} {Phys. Rev. Lett.}\ }\textbf {\bibinfo
  {volume} {8}},\ \bibinfo {pages} {428} (\bibinfo {year} {1962})}\BibitemShut
  {NoStop}%
\bibitem [{\citenamefont {Moriya}(1956)}]{Moriya1956}%
  \BibitemOpen
  \bibfield  {author} {\bibinfo {author} {\bibfnamefont {T.}~\bibnamefont
  {Moriya}},\ }\href@noop {} {\bibfield  {journal} {\bibinfo  {journal}
  {Progress of Theoretical Physics (Kyoto)}\ }\textbf {\bibinfo {volume}
  {16}},\ \bibinfo {pages} {641} (\bibinfo {year} {1956})}\BibitemShut
  {NoStop}%
\bibitem [{\citenamefont {Jaccarino}(1965)}]{Jaccarino1965}%
  \BibitemOpen
  \bibfield  {author} {\bibinfo {author} {\bibfnamefont {V.}~\bibnamefont
  {Jaccarino}},\ }\href@noop {} {\emph {\bibinfo {title} {Nuclear Resonance in
  Antifewrromagnets}}},\ edited by\ \bibinfo {editor} {\bibfnamefont {G.~T.}\
  \bibnamefont {Rado}}\ and\ \bibinfo {editor} {\bibfnamefont {H.}~\bibnamefont
  {Suhl}},\ Vol.\ \bibinfo {volume} {Magnetism IIA}\ (\bibinfo  {publisher}
  {Academic Press},\ \bibinfo {year} {1965})\BibitemShut {NoStop}%
\bibitem [{\citenamefont {Chakravarty}\ \emph {et~al.}(1989)\citenamefont
  {Chakravarty}, \citenamefont {Halperin},\ and\ \citenamefont
  {Nelson}}]{Chakravarty}%
  \BibitemOpen
  \bibfield  {author} {\bibinfo {author} {\bibfnamefont {S.}~\bibnamefont
  {Chakravarty}}, \bibinfo {author} {\bibfnamefont {B.~I.}\ \bibnamefont
  {Halperin}}, \ and\ \bibinfo {author} {\bibfnamefont {D.~R.}\ \bibnamefont
  {Nelson}},\ }\href {\doibase 10.1103/PhysRevB.39.2344} {\bibfield  {journal}
  {\bibinfo  {journal} {Phys. Rev. B}\ }\textbf {\bibinfo {volume} {39}},\
  \bibinfo {pages} {2344} (\bibinfo {year} {1989})}\BibitemShut {NoStop}%
\bibitem [{\citenamefont {Pennington}\ and\ \citenamefont
  {Slichter}(1991)}]{PenningtonPRB1989}%
  \BibitemOpen
  \bibfield  {author} {\bibinfo {author} {\bibfnamefont {C.~H.}\ \bibnamefont
  {Pennington}}\ and\ \bibinfo {author} {\bibfnamefont {C.~P.}\ \bibnamefont
  {Slichter}},\ }\href {\doibase 10.1103/PhysRevLett.66.381} {\bibfield
  {journal} {\bibinfo  {journal} {Phys. Rev. Lett.}\ }\textbf {\bibinfo
  {volume} {66}},\ \bibinfo {pages} {381} (\bibinfo {year} {1991})}\BibitemShut
  {NoStop}%
\bibitem [{\citenamefont {Pennington}\ \emph {et~al.}(1989)\citenamefont
  {Pennington}, \citenamefont {Durand}, \citenamefont {Slichter}, \citenamefont
  {Rice}, \citenamefont {Bukowski},\ and\ \citenamefont
  {Ginsberg}}]{PenningtonPRL1991}%
  \BibitemOpen
  \bibfield  {author} {\bibinfo {author} {\bibfnamefont {C.~H.}\ \bibnamefont
  {Pennington}}, \bibinfo {author} {\bibfnamefont {D.~J.}\ \bibnamefont
  {Durand}}, \bibinfo {author} {\bibfnamefont {C.~P.}\ \bibnamefont
  {Slichter}}, \bibinfo {author} {\bibfnamefont {J.~P.}\ \bibnamefont {Rice}},
  \bibinfo {author} {\bibfnamefont {E.~D.}\ \bibnamefont {Bukowski}}, \ and\
  \bibinfo {author} {\bibfnamefont {D.~M.}\ \bibnamefont {Ginsberg}},\ }\href
  {\doibase 10.1103/PhysRevB.39.274} {\bibfield  {journal} {\bibinfo  {journal}
  {Phys. Rev. B}\ }\textbf {\bibinfo {volume} {39}},\ \bibinfo {pages} {274}
  (\bibinfo {year} {1989})}\BibitemShut {NoStop}%
\bibitem [{\citenamefont {Imai}\ \emph
  {et~al.}(1993{\natexlab{c}})\citenamefont {Imai}, \citenamefont {Slichter},
  \citenamefont {Paulikas},\ and\ \citenamefont {Veal}}]{ImaiPRB1993}%
  \BibitemOpen
  \bibfield  {author} {\bibinfo {author} {\bibfnamefont {T.}~\bibnamefont
  {Imai}}, \bibinfo {author} {\bibfnamefont {C.~P.}\ \bibnamefont {Slichter}},
  \bibinfo {author} {\bibfnamefont {A.~P.}\ \bibnamefont {Paulikas}}, \ and\
  \bibinfo {author} {\bibfnamefont {B.}~\bibnamefont {Veal}},\ }\href {\doibase
  10.1103/PhysRevB.47.9158} {\bibfield  {journal} {\bibinfo  {journal} {Phys.
  Rev. B}\ }\textbf {\bibinfo {volume} {47}},\ \bibinfo {pages} {9158}
  (\bibinfo {year} {1993}{\natexlab{c}})}\BibitemShut {NoStop}%
\bibitem [{\citenamefont {Itoh}\ \emph {et~al.}(1992)\citenamefont {Itoh},
  \citenamefont {Yasuoka}, \citenamefont {Fujiwara}, \citenamefont {Ueda},
  \citenamefont {Machi}, \citenamefont {Tomeno}, \citenamefont {Tai},
  \citenamefont {Koshizuka},\ and\ \citenamefont {Tanaka}}]{ItohJPSJ1992}%
  \BibitemOpen
  \bibfield  {author} {\bibinfo {author} {\bibfnamefont {Y.}~\bibnamefont
  {Itoh}}, \bibinfo {author} {\bibfnamefont {H.}~\bibnamefont {Yasuoka}},
  \bibinfo {author} {\bibfnamefont {Y.}~\bibnamefont {Fujiwara}}, \bibinfo
  {author} {\bibfnamefont {Y.}~\bibnamefont {Ueda}}, \bibinfo {author}
  {\bibfnamefont {T.}~\bibnamefont {Machi}}, \bibinfo {author} {\bibfnamefont
  {I.}~\bibnamefont {Tomeno}}, \bibinfo {author} {\bibfnamefont
  {K.}~\bibnamefont {Tai}}, \bibinfo {author} {\bibfnamefont {N.}~\bibnamefont
  {Koshizuka}}, \ and\ \bibinfo {author} {\bibfnamefont {S.}~\bibnamefont
  {Tanaka}},\ }\href@noop {} {\bibfield  {journal} {\bibinfo  {journal} {J.
  Phys. Soc. Jpn.}\ }\textbf {\bibinfo {volume} {61}},\ \bibinfo {pages} {1287}
  (\bibinfo {year} {1992})}\BibitemShut {NoStop}%
\bibitem [{\citenamefont {Takigawa}(1994)}]{TakigawaPRB1994}%
  \BibitemOpen
  \bibfield  {author} {\bibinfo {author} {\bibfnamefont {M.}~\bibnamefont
  {Takigawa}},\ }\href {\doibase 10.1103/PhysRevB.49.4158} {\bibfield
  {journal} {\bibinfo  {journal} {Phys. Rev. B}\ }\textbf {\bibinfo {volume}
  {49}},\ \bibinfo {pages} {4158} (\bibinfo {year} {1994})}\BibitemShut
  {NoStop}%
\bibitem [{Rel()}]{Relation}%
  \BibitemOpen
  \href@noop {} {\emph {\bibinfo {title} {$1/T_{2G}$ measures only the
  so-called homogeneous linewidth caused by ``like'' nuclear-spins, whereas
  $\Delta f_{1/2}$ also reflects the the contributions from the ``unlike''
  nuclear spins and so-called inhomogeneous linewidth.}}}\BibitemShut {Stop}%
\bibitem [{\citenamefont {Tou}\ \emph {et~al.}(1992)\citenamefont {Tou},
  \citenamefont {Matsumura},\ and\ \citenamefont {Yamagata}}]{TouLBCOT2}%
  \BibitemOpen
  \bibfield  {author} {\bibinfo {author} {\bibfnamefont {H.}~\bibnamefont
  {Tou}}, \bibinfo {author} {\bibfnamefont {M.}~\bibnamefont {Matsumura}}, \
  and\ \bibinfo {author} {\bibfnamefont {H.}~\bibnamefont {Yamagata}},\
  }\href@noop {} {\bibfield  {journal} {\bibinfo  {journal} {J. Phys. Soc.
  Jpn.}\ }\textbf {\bibinfo {volume} {61}},\ \bibinfo {pages} {1477} (\bibinfo
  {year} {1992})}\BibitemShut {NoStop}%
\bibitem [{\citenamefont {Castro~Neto}\ and\ \citenamefont
  {Hone}(1996)}]{CastronetoPRL76}%
  \BibitemOpen
  \bibfield  {author} {\bibinfo {author} {\bibfnamefont {A.~H.}\ \bibnamefont
  {Castro~Neto}}\ and\ \bibinfo {author} {\bibfnamefont {D.}~\bibnamefont
  {Hone}},\ }\href {\doibase 10.1103/PhysRevLett.76.2165} {\bibfield  {journal}
  {\bibinfo  {journal} {Phys. Rev. Lett.}\ }\textbf {\bibinfo {volume} {76}},\
  \bibinfo {pages} {2165} (\bibinfo {year} {1996})}\BibitemShut {NoStop}%
\bibitem [{\citenamefont {Sokol}\ and\ \citenamefont {Pines}(1993)}]{Sokol}%
  \BibitemOpen
  \bibfield  {author} {\bibinfo {author} {\bibfnamefont {A.}~\bibnamefont
  {Sokol}}\ and\ \bibinfo {author} {\bibfnamefont {D.}~\bibnamefont {Pines}},\
  }\href {\doibase 10.1103/PhysRevLett.71.2813} {\bibfield  {journal} {\bibinfo
   {journal} {Phys. Rev. Lett.}\ }\textbf {\bibinfo {volume} {71}},\ \bibinfo
  {pages} {2813} (\bibinfo {year} {1993})}\BibitemShut {NoStop}%
\bibitem [{\citenamefont {Thurber}\ \emph {et~al.}(2000)\citenamefont
  {Thurber}, \citenamefont {Imai}, \citenamefont {Saitoh}, \citenamefont
  {Azuma}, \citenamefont {Takano},\ and\ \citenamefont {Chou}}]{Thurber3leg}%
  \BibitemOpen
  \bibfield  {author} {\bibinfo {author} {\bibfnamefont {K.~R.}\ \bibnamefont
  {Thurber}}, \bibinfo {author} {\bibfnamefont {T.}~\bibnamefont {Imai}},
  \bibinfo {author} {\bibfnamefont {T.}~\bibnamefont {Saitoh}}, \bibinfo
  {author} {\bibfnamefont {M.}~\bibnamefont {Azuma}}, \bibinfo {author}
  {\bibfnamefont {M.}~\bibnamefont {Takano}}, \ and\ \bibinfo {author}
  {\bibfnamefont {F.~C.}\ \bibnamefont {Chou}},\ }\href {\doibase
  10.1103/PhysRevLett.84.558} {\bibfield  {journal} {\bibinfo  {journal} {Phys.
  Rev. Lett.}\ }\textbf {\bibinfo {volume} {84}},\ \bibinfo {pages} {558}
  (\bibinfo {year} {2000})}\BibitemShut {NoStop}%
\bibitem [{\citenamefont {Imai~et al.}(shed)}]{Imai_inhomogeneity}%
  \BibitemOpen
  \bibfield  {author} {\bibinfo {author} {\bibfnamefont {T.}~\bibnamefont
  {Imai~et al.}},\ }\href@noop {} {} (\bibinfo {year} {To be
  published})\BibitemShut {NoStop}%
\bibitem [{\citenamefont {Arguello}\ \emph {et~al.}(2014)\citenamefont
  {Arguello}, \citenamefont {Chockalingam}, \citenamefont {Rosenthal},
  \citenamefont {Zhao}, \citenamefont {Guti\'errez}, \citenamefont {Kang},
  \citenamefont {Chung}, \citenamefont {Fernandes}, \citenamefont {Jia},
  \citenamefont {Millis}, \citenamefont {Cava},\ and\ \citenamefont
  {Pasupathy}}]{STM}%
  \BibitemOpen
  \bibfield  {author} {\bibinfo {author} {\bibfnamefont {C.~J.}\ \bibnamefont
  {Arguello}}, \bibinfo {author} {\bibfnamefont {S.~P.}\ \bibnamefont
  {Chockalingam}}, \bibinfo {author} {\bibfnamefont {E.~P.}\ \bibnamefont
  {Rosenthal}}, \bibinfo {author} {\bibfnamefont {L.}~\bibnamefont {Zhao}},
  \bibinfo {author} {\bibfnamefont {C.}~\bibnamefont {Guti\'errez}}, \bibinfo
  {author} {\bibfnamefont {J.~H.}\ \bibnamefont {Kang}}, \bibinfo {author}
  {\bibfnamefont {W.~C.}\ \bibnamefont {Chung}}, \bibinfo {author}
  {\bibfnamefont {R.~M.}\ \bibnamefont {Fernandes}}, \bibinfo {author}
  {\bibfnamefont {S.}~\bibnamefont {Jia}}, \bibinfo {author} {\bibfnamefont
  {A.~J.}\ \bibnamefont {Millis}}, \bibinfo {author} {\bibfnamefont {R.~J.}\
  \bibnamefont {Cava}}, \ and\ \bibinfo {author} {\bibfnamefont {A.~N.}\
  \bibnamefont {Pasupathy}},\ }\href {\doibase 10.1103/PhysRevB.89.235115}
  {\bibfield  {journal} {\bibinfo  {journal} {Phys. Rev. B}\ }\textbf {\bibinfo
  {volume} {89}},\ \bibinfo {pages} {235115} (\bibinfo {year}
  {2014})}\BibitemShut {NoStop}%
\bibitem [{\citenamefont {Kivelson}\ \emph {et~al.}(1998)\citenamefont
  {Kivelson}, \citenamefont {Fradkin},\ and\ \citenamefont {Emery}}]{Kivelson}%
  \BibitemOpen
  \bibfield  {author} {\bibinfo {author} {\bibfnamefont {S.~A.}\ \bibnamefont
  {Kivelson}}, \bibinfo {author} {\bibfnamefont {E.}~\bibnamefont {Fradkin}}, \
  and\ \bibinfo {author} {\bibfnamefont {V.~J.}\ \bibnamefont {Emery}},\
  }\href@noop {} {\bibfield  {journal} {\bibinfo  {journal} {Nature}\ }\textbf
  {\bibinfo {volume} {393}},\ \bibinfo {pages} {550} (\bibinfo {year}
  {1998})}\BibitemShut {NoStop}%
\bibitem [{\citenamefont {Fu}\ \emph {et~al.}(2012)\citenamefont {Fu},
  \citenamefont {Torchetti}, \citenamefont {Imai}, \citenamefont {Ning},
  \citenamefont {Yan},\ and\ \citenamefont {Sefat}}]{Fu1111PRL}%
  \BibitemOpen
  \bibfield  {author} {\bibinfo {author} {\bibfnamefont {M.}~\bibnamefont
  {Fu}}, \bibinfo {author} {\bibfnamefont {D.~A.}\ \bibnamefont {Torchetti}},
  \bibinfo {author} {\bibfnamefont {T.}~\bibnamefont {Imai}}, \bibinfo {author}
  {\bibfnamefont {F.~L.}\ \bibnamefont {Ning}}, \bibinfo {author}
  {\bibfnamefont {J.-Q.}\ \bibnamefont {Yan}}, \ and\ \bibinfo {author}
  {\bibfnamefont {A.~S.}\ \bibnamefont {Sefat}},\ }\href {\doibase
  10.1103/PhysRevLett.109.247001} {\bibfield  {journal} {\bibinfo  {journal}
  {Phys. Rev. Lett.}\ }\textbf {\bibinfo {volume} {109}},\ \bibinfo {pages}
  {247001} (\bibinfo {year} {2012})}\BibitemShut {NoStop}%
\bibitem [{\citenamefont {Tou}\ \emph {et~al.}(1993)\citenamefont {Tou},
  \citenamefont {Matsumura},\ and\ \citenamefont {Yamagata}}]{TouLBCOT1}%
  \BibitemOpen
  \bibfield  {author} {\bibinfo {author} {\bibfnamefont {H.}~\bibnamefont
  {Tou}}, \bibinfo {author} {\bibfnamefont {M.}~\bibnamefont {Matsumura}}, \
  and\ \bibinfo {author} {\bibfnamefont {H.}~\bibnamefont {Yamagata}},\
  }\href@noop {} {\bibfield  {journal} {\bibinfo  {journal} {J. Phys. Soc.
  Jpn.}\ }\textbf {\bibinfo {volume} {62}},\ \bibinfo {pages} {1474} (\bibinfo
  {year} {1993})}\BibitemShut {NoStop}%
\bibitem [{\citenamefont {Pelc}\ \emph {et~al.}(2017)\citenamefont {Pelc},
  \citenamefont {Grafe}, \citenamefont {Gu},\ and\ \citenamefont
  {Po\ifmmode~\check{z}\else \v{z}\fi{}ek}}]{Pelc}%
  \BibitemOpen
  \bibfield  {author} {\bibinfo {author} {\bibfnamefont {D.}~\bibnamefont
  {Pelc}}, \bibinfo {author} {\bibfnamefont {H.-J.}\ \bibnamefont {Grafe}},
  \bibinfo {author} {\bibfnamefont {G.~D.}\ \bibnamefont {Gu}}, \ and\ \bibinfo
  {author} {\bibfnamefont {M.}~\bibnamefont {Po\ifmmode~\check{z}\else
  \v{z}\fi{}ek}},\ }\href {\doibase 10.1103/PhysRevB.95.054508} {\bibfield
  {journal} {\bibinfo  {journal} {Phys. Rev. B}\ }\textbf {\bibinfo {volume}
  {95}},\ \bibinfo {pages} {054508} (\bibinfo {year} {2017})}\BibitemShut
  {NoStop}%
\bibitem [{\citenamefont {Baek}\ \emph {et~al.}(2015)\citenamefont {Baek},
  \citenamefont {Utz}, \citenamefont {H\"ucker}, \citenamefont {Gu},
  \citenamefont {B\"uchner},\ and\ \citenamefont {Grafe}}]{BaekLaT1PRB2015}%
  \BibitemOpen
  \bibfield  {author} {\bibinfo {author} {\bibfnamefont {S.-H.}\ \bibnamefont
  {Baek}}, \bibinfo {author} {\bibfnamefont {Y.}~\bibnamefont {Utz}}, \bibinfo
  {author} {\bibfnamefont {M.}~\bibnamefont {H\"ucker}}, \bibinfo {author}
  {\bibfnamefont {G.~D.}\ \bibnamefont {Gu}}, \bibinfo {author} {\bibfnamefont
  {B.}~\bibnamefont {B\"uchner}}, \ and\ \bibinfo {author} {\bibfnamefont
  {H.-J.}\ \bibnamefont {Grafe}},\ }\href {\doibase 10.1103/PhysRevB.92.155144}
  {\bibfield  {journal} {\bibinfo  {journal} {Phys. Rev. B}\ }\textbf {\bibinfo
  {volume} {92}},\ \bibinfo {pages} {155144} (\bibinfo {year}
  {2015})}\BibitemShut {NoStop}%
\bibitem [{\citenamefont {Komiya}\ and\ \citenamefont {Ando}(2004)}]{Ando}%
  \BibitemOpen
  \bibfield  {author} {\bibinfo {author} {\bibfnamefont {S.}~\bibnamefont
  {Komiya}}\ and\ \bibinfo {author} {\bibfnamefont {Y.}~\bibnamefont {Ando}},\
  }\href {\doibase 10.1103/PhysRevB.70.060503} {\bibfield  {journal} {\bibinfo
  {journal} {Phys. Rev. B}\ }\textbf {\bibinfo {volume} {70}},\ \bibinfo
  {pages} {060503} (\bibinfo {year} {2004})}\BibitemShut {NoStop}%
\bibitem [{\citenamefont {Thurber}\ \emph {et~al.}(1997)\citenamefont
  {Thurber}, \citenamefont {Hunt}, \citenamefont {Imai}, \citenamefont {Chou},\
  and\ \citenamefont {Lee}}]{Thurber2122}%
  \BibitemOpen
  \bibfield  {author} {\bibinfo {author} {\bibfnamefont {K.~R.}\ \bibnamefont
  {Thurber}}, \bibinfo {author} {\bibfnamefont {A.~W.}\ \bibnamefont {Hunt}},
  \bibinfo {author} {\bibfnamefont {T.}~\bibnamefont {Imai}}, \bibinfo {author}
  {\bibfnamefont {F.~C.}\ \bibnamefont {Chou}}, \ and\ \bibinfo {author}
  {\bibfnamefont {Y.~S.}\ \bibnamefont {Lee}},\ }\href {\doibase
  10.1103/PhysRevLett.79.171} {\bibfield  {journal} {\bibinfo  {journal} {Phys.
  Rev. Lett.}\ }\textbf {\bibinfo {volume} {79}},\ \bibinfo {pages} {171}
  (\bibinfo {year} {1997})}\BibitemShut {NoStop}%
\bibitem [{\citenamefont {Julien}\ \emph {et~al.}(1999)\citenamefont {Julien},
  \citenamefont {Borsa}, \citenamefont {Carretta}, \citenamefont
  {Horvati\ifmmode~\acute{c}\else \'{c}\fi{}}, \citenamefont {Berthier},\ and\
  \citenamefont {Lin}}]{Julien1999PRL}%
  \BibitemOpen
  \bibfield  {author} {\bibinfo {author} {\bibfnamefont {M.-H.}\ \bibnamefont
  {Julien}}, \bibinfo {author} {\bibfnamefont {F.}~\bibnamefont {Borsa}},
  \bibinfo {author} {\bibfnamefont {P.}~\bibnamefont {Carretta}}, \bibinfo
  {author} {\bibfnamefont {M.}~\bibnamefont {Horvati\ifmmode~\acute{c}\else
  \'{c}\fi{}}}, \bibinfo {author} {\bibfnamefont {C.}~\bibnamefont {Berthier}},
  \ and\ \bibinfo {author} {\bibfnamefont {C.~T.}\ \bibnamefont {Lin}},\ }\href
  {\doibase 10.1103/PhysRevLett.83.604} {\bibfield  {journal} {\bibinfo
  {journal} {Phys. Rev. Lett.}\ }\textbf {\bibinfo {volume} {83}},\ \bibinfo
  {pages} {604} (\bibinfo {year} {1999})}\BibitemShut {NoStop}%
\bibitem [{\citenamefont {Narath}(1967)}]{Narath}%
  \BibitemOpen
  \bibfield  {author} {\bibinfo {author} {\bibfnamefont {A.}~\bibnamefont
  {Narath}},\ }\href {\doibase 10.1103/PhysRev.162.320} {\bibfield  {journal}
  {\bibinfo  {journal} {Phys. Rev.}\ }\textbf {\bibinfo {volume} {162}},\
  \bibinfo {pages} {320} (\bibinfo {year} {1967})}\BibitemShut {NoStop}%
\end{thebibliography}

%

\end{document}